\documentclass[english,12pt]{article}
\usepackage[latin1]{inputenc}
\usepackage{babel}
\usepackage{color}

\usepackage{amsxtra}
\usepackage{stmaryrd}

\usepackage{amsfonts}
\usepackage{amsthm}
\usepackage{amsmath}
\usepackage{amssymb}
\usepackage{bbm}
\usepackage{hyperref}

\usepackage{dsfont}

\newtheorem{thm}{Theorem}[section]
\newtheorem{defi}[thm]{Definition}
\newtheorem{prop}[thm]{Proposition}
\newtheorem{cor}[thm]{Corollary}
\newtheorem{lemma}[thm]{Lemma}

\theoremstyle{definition}
\newtheorem{remark}[thm]{Remark}
\newtheorem{example}[thm]{Example}

\newtheorem*{ak}{Acknowledgement}

\newcommand{\bt}{\begin{thm}}
\newcommand{\et}{\end{thm}}
\newcommand{\br}{\begin{remark}}
\newcommand{\er}{\end{remark}}
\newcommand{\bl}{\begin{lemma}}
\newcommand{\el}{\end{lemma}}
\newcommand{\bp}{\begin{proof}}
\newcommand{\ep}{\end{proof}}
\newcommand{\bal}{\begin{align*}}
\newcommand{\eal}{\end{align*}}
\newcommand{\bi}{\begin{itemize}}
\newcommand{\be}{\begin{equation}}
\newcommand{\ee}{\end{equation}}
\newcommand{\bea}{\begin{eqnarray}}
\newcommand{\eea}{\end{eqnarray}}
\newcommand{\ba}{\begin{align*}}
\newcommand{\ea}{\end{align*}}
\newcommand{\ei}{\end{itemize}}

\DeclareMathOperator{\Var}{Var}
\DeclareMathOperator{\Cov}{Cov}

\newcommand{\R}{\mathbb{R}}

\newcommand{\N}{\mathbb{N}}

\newcommand{\F}{\mathcal{F}}
\newcommand{\cF}{\mathcal{F}}

\newcommand{\FF}{\mathbb{F}}

\newcommand{\Om}{\Omega}
\newcommand{\om}{\omega}
\newcommand{\T}{\top}

\newcommand{\LiiP}{L^2(P)}

\newcommand{\cH}{\mathcal{H}}

\newcommand{\cM}{\mathcal{M}}

\newcommand{\cP}{\mathcal{P}}

\newcommand{\vt}{\vartheta}
\newcommand{\hvt}{\widehat\vartheta}
\newcommand{\hvp}{\widehat\varphi}
\newcommand{\tvt}{\widetilde\vartheta}
\newcommand{\tvp}{\widetilde\varphi}
\newcommand{\vp}{\varphi}

\newcommand{\ve}{\varepsilon}
\newcommand{\Ker}{\mathrm{Ker}}

\newcommand{\TS}{\Theta}
\newcommand{\omt}{(\om,t)}
\newcommand{\OmT}{\Om\times[0,T]}

\newcommand{\PtB}{P\otimes B}

\newcommand{\la}{\langle}
\newcommand{\ra}{\rangle}
\newcommand{\E}{\mathcal{E}}

\newcommand{\sint}{\stackrel{\mbox{\tiny$\bullet$}}{}}

\def\one{\mathbbm{1}}
\numberwithin{equation}{section}
\begin{document}

\title{\vspace{-0.4cm}Time-Consistent Mean-Variance Portfolio Selection in Discrete and Continuous Time}
\author{\vspace{-0.2cm}
Christoph Czichowsky\\
\\
Faculty of Mathematics, University of Vienna \\
Nordbergstrasse 15 \\
A-1090 Vienna, Austria \\
\vspace{-0.2cm}\\
{\tt christoph.czichowsky@univie.ac.at}
\\
\vspace{-0.2cm}\\
This version: March 16, 2012.\footnote{To appear in Finance \& Stochastics.}\\
First version: September 14, 2010.\footnote{NCCR FINRISK working paper No. 661.\hfill\break
Available at \href{http://www.nccr-finrisk.uzh.ch/media/pdf/wp/WP661_D1.pdf}{\tt http://www.nccr-finrisk.uzh.ch/media/pdf/wp/WP661\_D1.pdf}
}
}
\date{}
\maketitle
\vspace{-0.6cm}
\begin{abstract}
\noindent
It is well known that mean-variance portfolio selection is a time-inconsistent optimal control problem in the sense that it does not satisfy Bellman's optimality principle and therefore the usual dynamic programming approach fails. We develop a time-consistent formulation of this problem, which is based on a local notion of optimality called local mean-variance efficiency, in a general semimartingale setting. We start in discrete time, where the formulation is straightforward, and then find the natural extension to continuous time.  This complements and generalises the formulation by Basak and Chabakauri (2010) and the corresponding example in Bj\"ork and Murgoci (2010), where the treatment and the notion of optimality rely on an underlying Markovian framework. We justify the continuous-time formulation by showing that it coincides with the continuous-time limit of the discrete-time formulation. The proof of this convergence is based on a global description of the locally optimal strategy in terms of the structure condition and the F\"ollmer--Schweizer decomposition of the mean-variance tradeoff. As a byproduct, this also gives new convergence results for the F\"ollmer--Schweizer decomposition, i.e.~for locally risk minimising strategies.
\end{abstract}
\vspace{-0.2cm}\newline
\vspace{-0.2cm}\newline
\section{Introduction}
In his seminal paper ``Portfolio selection'' \cite{Markowitz}, Harry Markowitz gave to the common wisdom that investors try to maximise return and minimise risk a quantitative description by saying that the return should be measured by the expectation and the risk by the variance. In a one-period financial market, \emph{mean-variance portfolio selection} then simply consists of finding the self-financing portfolio whose one-period terminal wealth has maximal mean and minimal variance. Since the mean-variance criterion is quadratic with respect to the strategy, one can calculate the solution, the so-called \emph{mean-variance efficient strategy}, directly and explicitly. Apart from the appealing and immediate interpretation of the optimisation criterion this probably explains its popularity.

Although one can obtain explicit formulas in one period, a multiperiod or continuous-time treatment is considerably more delicate; this has already been observed by Mossin in \cite{Mossin}. The reason is the well-known fact that the mean-variance criterion does not satisfy Bellman's optimality principle.

One way to deal with this issue is to treat mean-variance portfolio selection as in the \emph{Markowitz problem} considered by Richardson \cite{R89}, Schweizer \cite{S94} and Li and Ng \cite{LiNg}. It consists of simply plugging in the multiperiod or continuous-time terminal wealth into the one period criterion and to maximise that with respect to the strategy over the entire time interval. Although this formulation fails to produce a time-consistent solution in the sense that it is optimal for the conditional criterion at a later time, this is nevertheless a common way to avoid dealing with time inconsistency of the mean-variance criterion used in the literature. There it is sometimes referred to as mean-variance portfolio selection under precommitment, as the investor commits to follow the strategy which is optimal at time zero even though it is not (conditionally) optimal later on.

 Alternatively, one can optimise the conditional mean-variance criterion myopically in each step over the gains in the next period as in Section 2.1.1 of \cite{CV02} in discrete time for example. Due to the myopic way of optimisation we call this strategy \emph{myopically mean-variance efficient} in this paper. For a continuous-time formulation of this one has then to pass to a limit in an appropriate way. Under the name local utility maximisation such a limit formulation has been developed in \cite{K99} and~\cite{K02} by Kallsen for utility maximisation problems.
 
In this paper, we approach the time inconsistency of the mean-variance criterion in a different way. We try to find a solution which is in some reasonable way optimal for the conditional mean-variance criterion and time-consistent in the sense that if it is optimal at time zero, it is also optimal on any remaining time interval. In a Markovian framework, such a time-consistent formulation has been introduced by Basak and Chabakauri in \cite{BC}. However, to find a time-consistent formulation in general is an open problem as pointed out by Schweizer at the end of the survey article \cite{S09}. As the failure of Bellman's optimality principle indicates, we have to use a different notion of optimality for the dynamic criterion than the classical one used in dynamic programming. As in \cite{BC}, we follow Robert Strotz who suggested in \cite{S56} (for a different time-inconsistent deterministic optimisation problem) to maximise not over all possible future strategies, but only those one is actually going to follow. In discrete time, this leads to determining the optimal strategy by a backward recursion starting from the terminal date. For a continuous-time formulation one has to combine this \emph{recursive approach to time inconsistency} with a limit argument. In a Markovian framework, for optimal consumption problems with non-exponential discounting this has recently been studied by Ekeland and Lazrak in \cite{EL08a} and \cite{EL08b} and Ekeland and Pirvu in \cite{EP08a} and \cite{EP08b} and for mean-variance portfolio selection problems by Basak and Chabakauri \cite{BC} and Bj\"ork, Murgoci and Zhou \cite{BMZ}. These authors give the definition of the time-consistent solution via a backward recursion the interpretation of a Nash subgame perfect equilibrium strategy for an intrapersonal game. Building on these specific cases, Bj\"ork and Murgoci developed in \cite{BM} a ``general theory of Markovian time inconsistent stochastic control problems'' for various forms of time inconsistency in a Markovian setting. In all these problems, however, one exploits that the underlying Markovian structure turns all quantities of interest into deterministic functions. Then recursive optimality can be characterised by a system of partial differential equations (PDEs), so-called extended Hamilton--Jacobi--Bellman equations, and one can provide verification theorems which allow to deduce that if one has a smooth solution to the PDE, this gives the solution to the optimal control problem.

Although it is known how to formulate and handle time-inconsistent optimal control problems in a Markovian framework, it is an open question how to do this in a more general setting and how to apply martingale techniques to these kind of problems (see for example  page 54 in \cite{BM}). For the problem of mean-variance portfolio selection, we show how one can answer these questions in this paper.  Note, however, that we exploit the underlying linear-quadratic structure of the problem for this and only consider mean-variance portfolio selection here.  In discrete time, obtaining the time-consistent solution by recursive optimisation is straightforward. To find the natural extension of this formulation to continuous time, we introduce a local notion of optimality called \emph{local mean-variance efficiency}; this is a first main result and gives a mathematically precise formulation in a general semimartingale framework. In continuous time, the definition of local mean-variance efficiency is inspired by the concept of continuous-time local risk minimisation introduced by Schweizer in \cite{S88}. As we shall see, our formulation in discrete as well as in continuous time embeds time-consistent mean-variance portfolio selection in a natural way into the already existing quadratic optimisation problems in mathematical finance, i.e.~the Markowitz problem, mean-variance hedging, and local risk minimisation; see \cite{S01} and \cite{S09}. Moreover, we provide an alternative characterisation of the optimal strategy in terms of the structure condition and the F\"ollmer--Schweizer decomposition of the mean-variance tradeoff. This is a second main result and gives necessary and sufficient conditions for the existence of a solution. Besides this, we obtain an intuitive interpretation of the optimal strategy. On the one hand the investor maximises the conditional mean-variance criterion in a myopic way one step ahead  by choosing the myopically mean-variance efficient strategy. This generates a risk represented by the mean-variance tradeoff which he then minimises by local risk minimisation on the other hand. Using the alternative characterisation of the optimal strategy allows us to justify the continuous-time formulation by showing that it coincides with the continuous-time limit of the discrete-time formulation. This underlines that our reasoning in discrete time, where the solution is determined by a backward recursion, is consistent with the way of defining optimality in continuous time and is our third main result. On the technical side, the link to the F\"ollmer--Schweizer decomposition and local risk minimisation allows us to exploit and extend known results.

Time consistency also plays a central role in the formulation of forward dynamic utilities by Musiela and Zariphopoulou; see \cite{MZ03} and \cite{MZ06} for example. There it is used to characterise the dynamic evolution of utility random fields by the optimal portfolios via the martingale optimality principle. Conversely these optimal portfolios then satisfy Bellman's optimality principle for the corresponding forward dynamic utility by definition. In contrast to their approach we do not propose a conceptual way to generate time-consistent dynamic utility functions here but rather how to determine a time-consistent optimal strategy by means of local optimisation for the underlying conditional mean-variance preferences that are not time consistent.

Recently Cui et al.~proposed in \cite{Cetal} an alternative way to deal with the time inconsistency of the mean-variance criterion. Relaxing the self-financing condition by allowing the withdrawal of money out of the market, they obtain a strategy which dominates the solution for the Markowitz problem in the sense that while both strategies achieve the same mean-variance pair for the terminal wealth their optimal strategy enables the investor to receive a free cash flow stream during the investment process. Compared to our study their reasoning and techniques are different. In particular, their solution is not time consistent in our sense. 
 
The remainder of the article is organised as follows. In the next section we explain the basic problem and the issue of time inconsistency of the mean-variance criterion and introduce the required notation for this. To establish the time-consistent formulation, we start in Section \ref{sec:dt} in discrete time and then find the natural extension of that to continuous time in Section \ref{sec:ct}. The convergence of the solutions obtained in discretisations of a continuous-time model to the solution in continuous time is shown in the last section.
\section{Formulation of the problem and preliminaries}\label{se:fp}
Let $(\Omega,\mathcal{F},P)$ be a probability space with a filtration $\mathbb{F}=(\mathcal{F}_t)_{0\leq t\leq T}$ satisfying the usual conditions of completeness and right-continuity, where $T\in(0,\infty)$ is a fixed and finite time horizon. For all unexplained notation concerning stochastic integration we refer to the book of Dellacherie and Meyer \cite{DM82}. Our presentation of the basic problem here builds upon that in Basak and Chabakauri \cite{BC} and Schweizer \cite{S09}.

We consider a \emph{financial market} consisting of one riskless asset whose price is $1$ and $d$ risky assets described by an $\R^d$-valued semimartingale $S$. As set of \emph{trading strategies} we choose $\TS:=\Theta_{S}:=\{\vt\in L(S)~|~\int \vt dS \in\cH^2(P)\}$ where $L(S)$ is the space of all $\R^d$-valued, $S$-integrable, predictable processes and $\cH^2(P)$ the space of all square-integrable semimartingales, i.e.~special semimartingales $X$ with canonical decomposition \mbox{$X=X_0+M^X+A^X$} such that
$$\textstyle\| X\|_{\cH^2(P)}:=\|X_0\|_{\LiiP}+\big\|\big([M^X,M^X]_T\big)^{\frac{1}{2}}\big\|_{\LiiP}+\big\|\int_0^T|d A^X_s|\big\|_{\LiiP}<+\infty.$$
The \emph{wealth} generated by using the self-financing trading strategy $\vt\in\Theta$ up to time $t\in[0,T]$ and starting from initial capital $x\in\R$ is given by
$$\textstyle V_t(x,\vt):=x+\int_0^t\vt_u dS_u=:x+\vt\sint S_t.$$
Note that we use the notation above also for the stochastic integral in discrete time. Since we work with $\TS_S$, we can always find \emph{representative square-integrable portfolios} for the financial market $(S,\TS_S)$ as explained in the appendix. These are portfolios $\vp^i\in\TS_S$ for $i=1,\ldots,d$ such that the financial market $(\widetilde S, \Theta_{\widetilde S})$ with $\widetilde S^i:=\vp^ i\sint S$ for $i=1,\ldots,d$ satisfies $\widetilde S\in\cH^2(P)$ and which are representative in the sense that $(\widetilde S, \Theta_{\widetilde S})$ generates the same wealth processes as $(S,\TS_S)$, i.e.~\mbox{$\TS_S\sint S=\Theta_{\widetilde S}\sint {\widetilde S}$}. We can and do therefore assume without loss of generality that $S$ is in $\mathcal{H}^2(P)$ and hence special with canonical decomposition $S=S_0+M+A$, where $M$ is an $\R^d$-valued square-integrable martingale null at zero, i.e.~$M\in\cM^2_{0}(P)$, and $A$ is an \mbox{$\R^d$-valued} predictable RCLL process, i.e.~right continuous with left limits (RCLL), null at zero with square-integrable variation. Besides simplifying the presentation this allows to refer directly to the standard literature on quadratic optimisation in mathematical finance which usually assumes (local) square-integrability of $S$. Conversely, this change of parameterisation of the financial market can be used to generalise local risk minimisation and quadratic hedging to the case where $S$ is a general semimartingale and not necessarily locally square-integrable; this will be explained in more detail in future work.

In the one-period case, where $T=1$, $\vt\sint S_1=\vt^\T_1 (S_1-S_0)=:\vt^\T_1 \Delta S_1$ and $\vt_1$ is an $\F_0$-measurable $\R^d$-valued random vector, \emph{mean-variance portfolio selection (MVPS)} with risk aversion $\gamma>0$ can be formulated as the problem to
\be
\text{\textrm{maximise $E[x+\vt_1^\T\Delta S_1]-\frac{\gamma}{2}\Var[x+\vt^\T_1\Delta S_1]$ over all $\cF_0$-measurable $\vt_1$}}.\label{OPMVPS}
\ee
The solution, the so-called \emph{mean-variance efficient strategy}, is then
\be
\tvt_1:=\frac{1}{\gamma}\Cov[\Delta S_1|\cF_0]^{-1}E[\Delta S_1|\cF_0]=:\hvt_1\label{opmvest}
\ee
which is given by an explicit formula in terms of the risk aversion and the conditional mean and variance of the stock price changes. Note that $\Cov[\Delta S_1|\cF_0]^{-1}$ denotes the Moore-Penrose pseudoinverse (see \cite{A72} for example) and therefore the solution exists if and only if $E[\Delta S_1|\cF_0]$ is in the range of $\Cov[\Delta S_1|\cF_0]$.

Having obtained the formulation and the explicit form of the solution in one period, we ask how the two extend to multiperiod or continuous time. An immediate extension of the formulation is simply to plug in the multiperiod or continuous-time terminal wealth into the one-period criterion. This corresponds to considering mean-variance portfolio selection (MVPS) as the  problem to
\be
\text{\textrm{maximise $E[V_T(x,\vartheta)]-\frac{\gamma}{2}\Var[V_T(x,\vartheta)]$ over all $\vt\in\Theta$}}.\label{MVPS}
\ee
The latter is an alternative formulation of the classical \emph{Markowitz problem} to
\begin{align}
&\text{minimise $\Var[V_T(x,\vartheta)]=E\big[|V_T(x,\vartheta)|^2\big]-m^2$}\nonumber\\
&\text{subject to $E[V_T(x,\vartheta)]=m>x$ and $\vartheta\in\TS$.}\label{MP}
\end{align}

In this set-up, MVPS is a \emph{static optimisation problem} as one determines the optimal strategy $\tvt$ for \eqref{MVPS} over the entire time interval with respect to the criterion evaluated at time $0$.  To obtain the solutions to \eqref{MVPS} and \eqref{MP} it can be shown by elementary Hilbert space arguments (see for example \cite{SW06}) that these are related to the solution of an auxiliary problem: If $1-\widetilde\vp\sint S_T\not\equiv0$ and $E[\widetilde\vp\sint S_T]\ne0$, the solutions $\tvt$ and $\tvt^{(x,m)}$ to \eqref{MVPS} and \eqref{MP} are given by
\be
\text{$\tvt=\frac{1}{\gamma}\frac{1}{E[1-\widetilde\vp\sint S_T]}\widetilde\vp\quad$and$\quad\tvt^{(m,x)}=\frac{m-x}{E[\widetilde\vp\sint S_T]}\widetilde\vp=(m-x)\gamma\frac{E[1-\widetilde\vp\sint S_T]}{E[\widetilde\vp\sint S_T]}\tvt,$}\label{gds}
\ee
where $\widetilde\vp$ is the solution to the auxiliary problem to
\be
\text{minimise $E\big[|V_T(-1,\vt)|^2\big]=E\big[|1-\vt\sint S_T|^2\big]$ over all $\vt\in\TS$}\label{ap-1}.
\ee
Since \eqref{ap-1} is a standard stochastic optimal control problem, it can be solved by dynamic programming and the dynamic structure of $\tvp$ can be described more explicitly, which via \eqref{gds} then gives a \emph{dynamic description} of $\tvt$ (and $\tvt^{(m,x)}$) as well. For this one considers instead of the single static problem \eqref{ap-1} the corresponding dynamic optimisation problem given by the conditional problems to
\be
\text{minimise $E\big[|V_T(-1,\vt)|^2\big|\F_t\big]=E\big[|1-\vt\sint S_T|^2\big|\F_t\big]$ over all $\vt\in\Theta_t(\psi)$}\label{Dap-1}
\ee
where $\Theta_t(\psi):=\{\vt\in\Theta~|~\vt\mathbbm{1}_{[\mskip-2mu[0,t]\mskip-2mu]}=\psi\mathbbm{1}_{[\mskip-2mu[0,t]\mskip-2mu]}\}$ denotes the set of all strategies $\vt\in\Theta$ that agree up to time $t$ with a given $\psi\in\Theta$. The family of conditional problems \eqref{Dap-1} is time consistent in the sense that it satisfies \emph{Bellman's optimality principle:} If $\tvp$ is the solution to \eqref{ap-1}, then it is for any $t\in[0,T]$ also optimal for the conditional criterion \eqref{Dap-1} with $\psi=\tvp$ on the remaining time interval $(t,T]$. This time consistency gives a dynamic characterisation of optimality of the solution $\tvp$ for the auxiliary problem \eqref{ap-1} via the dynamic optimisation problem \eqref{Dap-1} and is a conceptual aspect of the problem \eqref{ap-1}. Since the time consistency of the conditional problems \eqref{Dap-1} allows to compute the solution $\tvp$ recursively by dynamic programming, this indeed allows to describe $\tvp$ and hence also $\tvt$ via \eqref{gds} as dynamic processes on $[0,T]$ more explicitly; see \cite{S09} for references and a survey as well as \cite{CS11} for recent results obtained in a general semimartingale framework in this direction. For the solution $\tvt$ to the static MVPS problem \eqref{MVPS}, however, this is so far only a computational aspect.

To study \eqref{MVPS} as a \emph{dynamic optimisation problem}, a natural formulation is to consider in analogy to \eqref{Dap-1} the conditional problems to
\be
\text{\textrm{maximise $U_t(\vt):=E[V_T(x,\vartheta)|\cF_t]-\frac{\gamma}{2}\Var[V_T(x,\vartheta)|\cF_t]$ over all $\vt\in\TS_t(\psi)$}.}\label{DMVPS}
\ee
However, plugging in the optimal strategy $\tvt$ to \eqref{MVPS} for $\psi$ yields that, in contrast to \eqref{Dap-1}, this family of conditional problem is no longer time consistent and that Bellman's optimality principle fails: If we use the solution $\tvt$ to \eqref{MVPS} on $[0,t]$ and then determine the corresponding conditionally optimal strategy by maximising in \eqref{DMVPS} over all \mbox{$\vt\in\TS_t(\tvt)$}, then this strategy is different from $\tvt$ on $(t,T]$. This time inconsistency leads us to the basic question how to obtain a \emph{time-consistent} dynamic formulation of MVPS. That is to find a dynamic formulation that gives a solution $\hvt$ which is in some reasonable sense optimal for the time-inconsistent conditional mean-variance criterion $U_t(\cdot)$ at time $t$ for each $t\in[0,T]$. This is a conceptual problem. We remark that it depends of course on the preferences of the investor whether he would like to have a (so-called pre-commitment) strategy which involves dynamic trading and is optimal for the static mean-variance criterion \eqref{MVPS} evaluated at time $0$, or a strategy $\hvt$ which is optimal for the conditional mean-variance criterion in a dynamic and time-consistent sense.
The reason for the time inconsistency of the (conditional) mean-variance criterion in \eqref{DMVPS} is the conditional variance term. As explained in \cite{BC}, we see that due to the total variance formula
\begin{align*}
&\Var[V_T(x,\vartheta)|\cF_t]=E\big[\Var[V_T(x,\vartheta)|\cF_{t+h}]\big|\cF_t\big]+\Var\Big[E\big[\mbox{$\int_{t+h}^T\vt dS$}\big|\cF_{t+h}\big]+V_{t+h}(x,\vt)\Big|\cF_t\Big],
\end{align*}
the objective function at time $t$ is given by the conditional expectation of the objective function at time $t+h$ and some adjustment term, i.e.
\begin{align}
U_t(\vt)=E\big[U_{t+h}(\vt)\big|\cF_t\big]-\frac{\gamma}{2}\Var\Big[E\big[\mbox{$\int_{t+h}^T\vt dS$}\big|\cF_{t+h}\big]+V_{t+h}(x,\vt)\Big|\cF_t\Big]\label{dpe}
\end{align}
for all $\vt\in\Theta$. As this adjustment term does not only depend on the strategy via its behaviour on $(t,t+h]$ but also on $(t+h,T]$, it causes ``an incentive for the investor  to deviate from his optimal strategy at a later time'' as explained in \cite{BC}. Mathematically, the adjustment term cannot be interpreted as a running cost term, and therefore the objective function is not of the ``standard form'' which is crucial for the dynamic programming approach to work; see for instance \cite{BM}, or \cite{FS} for a textbook account. The economic explanation for the time-inconsistent behaviour of the investor is as follows. At time $t$, the investor uses the strategy on $(t+h,T]$ not only to maximise the time $(t+h)$ objective function $U_{t+h}(\vt)$, but also to minimise the second term. This means that he tries to minimise some of the risk coming from the strategy used on $(t,t+h]$. At time $t+h$, the outcome of the trading on $(t,t+h]$ is already known and there remains no risk to be minimised. Therefore the investor at time $t+h$ chooses the trading strategy on $(t+h,T]$ only to maximise $U_{t+h}(\vt)$, and so his objective and hence his choice will be in general different from that at time $t$.

An alternative explanation for the failure of the time consistency of the dynamic formulation \eqref{DMVPS} is of course that already the underlying mean-variance preferences are time inconsistent due to their non-monotonicity; see for example \cite{MMRT}.

Loosely speaking, the reason for the inconsistency of the formulation \eqref{DMVPS} is that we are optimising over too many strategies, as we are also considering strategies which we are not going to use later on. To overcome this, we follow the recursive approach to time inconsistency ``\ldots to choose the best [strategy] not among all available strategies, but among those one is actually going to follow.'' proposed by Strotz in \cite{S56} (for the deterministic optimal consumption problem with non-exponential discounting). The same reasoning also appears in the context of local risk minimisation introduced by Schweizer in \cite{S88} to deal with the time inconsistency of the formulation of global risk minimisation.  For a dynamic formulation of MVPS, this suggests that we have to weaken our optimality criterion and to optimise in \eqref{DMVPS} not globally on $(t,T]$, but only ``locally on an infinitesimally small time interval $(t,t+dt]$'' going backwards from $T$ and using the ``optimal strategy on $(t+dt,T]$''. Since the investor following this rule chooses for any $t\in[0,T]$ ``the strategy on $(t,t+dt]$'' that he has to determine at time $t$ optimally for his criterion $U_t(\cdot)$ at time $t$, he has no reason to deviate from this ``locally mean-variance optimal'' strategy for the dynamic optimisation problem, which therefore leads to a time-consistent behaviour. ``In some sense this formulation interpolates between dynamic optimisation for a fixed time horizon and step-by-step one period optimisation'' (as has been formulated by one of the referees). This way to address the time inconsistency of the mean-variance criterion has been developed by Basak and Chabakauri in \cite{BC} in a Markovian setting by using partial differential equations which are available in this framework. Since the concept of local optimisation in a general set-up is more intuitive and conceptually easier to understand in discrete time, we consider this case in the next section first before proceeding with the more delicate situation in continuous time.

\section{Discrete Time}\label{sec:dt}
In this section, we develop a time-consistent formulation for the mean-variance portfolio selection problem in discrete time and derive the general structure of the solution. As this mainly serves for the motivation of the continuous-time case, we restrict our presentation here for simplicity to the one dimensional case $d=1$.

Let $T\in\N$ and assume that trading only takes place at fixed times $k=0,1,\ldots,T$, where we choose at time $k$ the number of shares $\vt_{k+1}$ to be held over the time period $(k,k+1]$. In this setting, we obtain an optimal strategy by recursively optimising starting from $T$, which is equivalent to optimality with respect to local perturbations. This is then a time-consistent solution to MVPS in the recursively optimal sense introduced by Strotz in~\cite{S56}.  Since we are optimising the conditional criterion of the entire remaining time interval only with respect to the strategy used in the next time step as in the concept of local risk minimisation (see \cite{S88} for example), we call this notion of optimality local mean-variance efficiency due to the local nature of optimisation. Mathematically, this is then formulated as follows.

\begin{defi} Let $\psi\in\Theta$ be a strategy and $k\in\{1,\ldots, T\}$. A \emph{local perturbation of $\psi$ at time $k$} is any strategy $\vt\in\Theta$ with $\vt_j=\psi_j$ for all $j\ne k$. We call a trading strategy $\hvt\in\Theta$ \emph{locally mean-variance efficient (LMVE)} if
\be
U_{k-1}(\hvt)\geq U_{k-1}(\vt)\qquad\text{P-a.s.}\label{dd1}
\ee
for all $k=1,\ldots,T$ and any local perturbation $\vt\in\Theta$ of $\hvt$ at time $k$ or, equivalently,
\be
U_{k-1}(\hvt)\geq U_{k-1}(\hvt+\delta\mathbbm{1}_{\{k\}})\qquad\text{P-a.s.}\label{dd2}
\ee
for all $k=1,\ldots,T$ and any $\delta\in\Theta$.
\end{defi}
Note that since $U_t(\vt)=V_t(x,\vt)+U_t(\mathbbm{1}_{\rrbracket t,T\rrbracket }\vt)=:V_t(x,\vt)+\overline{U}_t(\vt)$, the structure of mean-variance preferences implies that conditions \eqref{dd1} and \eqref{dd2} do not depend for fixed $k$ on the strategy used on $\{0,\ldots, k-1\}$. This allows us to derive the following recursive formula for the LMVE strategy $\hvt$, which underlines the time-consistency of the solution  and also implies its uniqueness. This formula already appeared in a Markovian framework in Proposition 5 in \cite{BC} and in a discrete-time setting in an unpublished Master thesis by Sigrid K\"allblad.
\begin{lemma}\label{lrrd}
A strategy $\hvt\in\TS$ is LMVE if and only if it satisfies
\be
\widehat\vt_{k}=\frac{1}{\gamma}\frac{E[\Delta S_{k}|\F_{k-1}]}{\Var\left[\Delta S_{k}|\F_{k-1}\right]}-\frac{\Cov\left[\Delta S_{k},\sum_{i=k+1}^T\widehat\vt_{i}\Delta S_{i}|\F_{k-1}\right]}{\Var\left[\Delta S_{k}|\F_{k-1}\right]}\label{rrd}
\ee
for $k=1,\ldots,T$.
\end{lemma}
\bp
Plugging $\hvt$ and $\hvt+\delta\mathbbm{1}_{\{k\}}$ into \eqref{dpe}, we obtain that \eqref{dd2} is equivalent to
\begin{align}
-\delta_{k}\left(E[\Delta S_{k}|\cF_{k-1}]-\gamma\Cov\left[\Delta S_{k},\sum_{i=k}^T\widehat\vt_{i}\Delta S_{i}\bigg|\cF_{k-1}\right]\right)+\frac{\gamma}{2}\Var\left[\delta_{k}\Delta S_{k}|\cF_{k-1}\right]\geq 0\label{prrd}
\end{align}
for all $k=1,\ldots,T$ and any $\delta\in\Theta$. Since $\Var\left[\delta_{k}\Delta S_{k}|\cF_{k-1}\right]\geq 0$ for all $k=1,\ldots,T$ and any $\delta\in\Theta$, it follows immediately that $\hvt$ satisfies \eqref{dd2} if \eqref{rrd} holds.
For the converse, we argue by backward induction; so assume that \eqref{rrd} holds for $j=k+1, \ldots ,T$. Because the conditional covariance term in \eqref{prrd} vanishes on $\{\Var[\Delta S_{k}|\F_{k-1}]=0\}$, we set 
\mbox{$\ve=E[\Delta S_{k}|\cF_{k-1}]\mathbbm{1}_{\{\Var[\Delta S_{k}|\F_{k-1}]=0\}}$} and
$$\vp=\left(\frac{1}{\gamma}\frac{E[\Delta S_{k}|\F_{k-1}]}{\Var\big[\Delta S_{k}\big|\F_{k-1}\big]}-\frac{\Cov\big[\Delta S_{k},\sum_{i=k+1}^T\widehat\vt_{i}\Delta S_{i}\big|\F_{k-1}\big]}{\Var\big[\Delta S_{k}\big|\F_{k-1}\big]}-\hvt_k\right)\mathbbm{1}_{\{\Var[\Delta S_{k}|\cF_{k-1}] > 0\}}.$$
Then choosing $\delta=\ve\mathbbm{1}_{\{E[(\ve\Delta S_{k})^2|\F_{k-1}] \leq n\}}\mathbbm{1}_{\{k\}}\in\TS$ and $\delta=\vp\mathbbm{1}_{\{E[(\vp\Delta S_{k})^2|\F_{k-1}] \leq n\}}\mathbbm{1}_{\{k\}}\in\TS$
for each $n\in\N$ implies that $\ve=0$ and $\vp=0$, as we could otherwise derive a contradiction to \eqref{prrd}. By the Cauchy--Schwarz inequality and since $\ve=0$, the right-hand side of \eqref{rrd} is always well defined by setting $\frac{0}{0}=0$, and equal to $\hvt$ since $\vp=0$. This completes the proof.
\ep
To simplify \eqref{rrd}, we use the canonical decomposition of $S=S_0+ M + A$  into a martingale $M$ and a predictable process $A$, which is in discrete time given by the \emph{Doob decomposition}, i.e.~$M_0:=0=:A_0$, $\Delta A_k=E[\Delta S_k |\F_{k-1}]$ and $\Delta M_k=\Delta S_k-E[\Delta S_k |\F_{k-1}]$ for $k=1,\ldots,T$. Then 
\eqref{rrd} can be written as
\be
\hvt_k=\frac{1}{\gamma}\frac{\Delta A_{k}}{E\left[(\Delta  M_{k})^2|\F_{k-1}\right]}-\frac{\Cov\left[\Delta  M_{k},\sum_{i=k+1}^T\widehat\vt_{i}\Delta   A_{i}|\F_{k-1}\right]}{E\left[(\Delta  M_{k})^2|\F_{k-1}\right]}\label{rrddd}
\ee
for $k=1,\ldots,T$. From this it follows by the Cauchy--Schwarz inequality that the existence of a LMVE strategy $\hvt$ implies that $S$ satisfies the \emph{structure condition (SC)},~i.e. there exists a predictable process $\lambda$ given by
$$\lambda_k:=\frac{\Delta  A_{k}}{E\left[(\Delta  M_{k})^2|\F_{k-1}\right]}=\frac{E[\Delta S_{k}|\F_{k-1}]}{\Var\left[\Delta S_{k}|\F_{k-1}\right]}\qquad\text{for $k=1,\ldots,T$}$$
such that the \emph{mean-variance tradeoff (MVT) process}
$$K_k:=\sum_{i=1}^k \frac{\big(E[\Delta S_{i}|\F_{i-1}]\big)^2}{\Var\left[\Delta S_{i}|\F_{i-1}\right]}=\sum_{i=1}^k \lambda_i^2E\left[(\Delta  M_{i})^2|\F_{i-1}\right]=\sum_{i=1}^k \lambda_i\Delta  A_{i}\qquad\text{for $k=0,\ldots,T$}$$
is finite-valued. This is not surprising, as these quantities also appear naturally in other quadratic optimisation problems in mathematical finance; see \cite{S01}. For each $\vt\in\Theta$, we define the process of \emph{expected future gains} $Z(\vt)$ and the square integrable martingale $Y(\vt)$ of its canonical decomposition by
\begin{align*}
Z_k(\vt):&=E\left[\sum_{i=k+1}^T\vt_i\Delta S_i\bigg|\cF_k\right]=E\left[\sum_{i=k+1}^T\vt_i\Delta A_i\bigg|\cF_k\right]\\
&=E\left[\sum_{i=1}^T\vt_i\Delta A_i\bigg|\cF_k\right]-\sum_{i=1}^k\vt_i\Delta A_i\\
&=:Y_k(\vt)-\sum_{i=1}^k\vt_i\Delta A_i
\end{align*}
for $k=0,1,\ldots,T$. Note that for the LMVE strategy $\hvt$, the process $Z(\hvt)$ has already been introduced in a discrete-time setting in Sigrid K\"allblad's Master thesis and in the Markovian framework in \cite{BC} in discrete and continuous time, where it is a function \mbox{$Z_t(\hvt)=f(W_t,S_t,X_t,t)$} of time $t$ and the underlying state variables, i.e.~current wealth $W_t$, stock price $S_t$ and hidden Markov factor $X_t$. Using the \emph{Galtchouk--Kunita--Watanabe (GKW) decomposition}
$$\sum_{i=1}^T\vt_i\Delta A_i=Y_0(\vt)+\sum_{i=1}^T\xi_i(\vt)\Delta M_i+L_T(\vt)$$
of $Y(\vt)$ with a square-integrable martingale $L(\vt)$ strongly orthogonal to $M$, we can rewrite $Z(\vt)$ as
\be
Z_k(\vt)=Y_k(\vt)-\sum_{i=1}^k\vt_i\Delta A_i=Y_0(\vt)+\sum_{i=1}^k\xi_i(\vt)\Delta M_i+L_k(\vt)-\sum_{i=1}^k\vt_i\Delta A_i\label{dZd}
\ee
for $k=0,1,\ldots,T$. Inserting the last expression into \eqref{rrddd}, we can reformulate Lemma \ref{lrrd} by combining the above as follows.
\bl\label{lLMVEGKWd}
The LMVE strategy $\hvt$ exists if and only if we have both
\bi
\item[1)] $S$ satisfies (SC) with $\lambda\in L^2(M)$, i.e.~$K_T\in L^1(P)$.
\item[2)] There exists $\widehat\psi\in\Theta$ such that
\be
\widehat\psi=\frac{1}{\gamma}\lambda -\xi(\widehat\psi),\label{doptcond}
\ee
where $\xi(\widehat\psi)$ is the integrand in the GKW decomposition of $\sum_{i=1}^T\widehat\psi_i\Delta A_i$.
\ei
In that case, $\hvt=\widehat\psi$.
\el
\bp
By Lemma \ref{lrrd} the existence of a LMVE strategy $\hvt$ and a strategy satisfying \eqref{rrddd} are equivalent. As already explained, \eqref{rrddd} implies by the Cauchy--Schwarz inequality that $S$ satisfies (SC). Since we obtain
$$\Cov\left[\Delta  M_{k},\sum_{i=k+1}^T\widehat\vt_{i}\Delta   A_{i}\bigg|\F_{k-1}\right]=\Cov\left[\Delta  M_{k},Z_k(\hvt)\Big|\F_{k-1}\right]=\xi_k(\hvt)E\left[(\Delta  M_{k})^2|\F_{k-1}\right]$$
by simply plugging into \eqref{rrddd} the definition of $Z(\hvt)$ and \eqref{dZd}, it follows that $\hvt$ satisfies \eqref{doptcond} and, conversely, that each strategy $\widehat\psi\in\Theta_S$ satisfying \eqref{doptcond} is LMVE. Moreover, since $\hvt\in\TS=L^2(M)\cap L^2(A)$, we have that $Y_T(\hvt)=\sum_{i=1}^T\hvt_i\Delta A_i\in L^2(P)$ and therefore that $\xi(\hvt)\in L^2(M)$ by construction. Rewriting \eqref{doptcond}, this implies that $\lambda=\gamma\hvt+\xi(\hvt)$ is in $L^2(M)$ and $K_T\in L^1(P)$, which completes the proof.
\ep
Integrating both sides of \eqref{doptcond} with $\widehat\psi=\hvt$ with respect to $M$ and plugging in the GKW decomposition then gives
\begin{align*}
\sum_{i=1}^T\hvt_i\Delta  M_i&=\frac{1}{\gamma}\sum_{i=1}^T\lambda_i\Delta  M_i-\sum_{i=1}^T\xi_i(\hvt)\Delta  M_i\\
&=\frac{1}{\gamma}\sum_{i=1}^T\lambda_i\Delta  M_i+Y_0(\hvt)+L_T(\hvt)-\sum_{i=1}^T\hvt_i\Delta  A_i.
\end{align*}
After rearranging terms and adding $\frac{1}{\gamma} K_T=\frac{1}{\gamma}\sum_{i=1}^T \lambda_{i}\Delta A_i$ on both sides we arrive at
\be
\frac{1}{\gamma} K_T=Y_0(\hvt)+\sum_{i=1}^T\left(\frac{1}{\gamma} \lambda_{i}-\hvt_i\right)\Delta  M_i+\sum_{i=1}^T\left(\frac{1}{\gamma} \lambda_{i}-\hvt_i\right)\Delta  A_i+L_T(\hvt),\label{eq:acdt}
\ee
which means that the terminal value of the MVT process $K_T$ admits a decomposition
\be
K_T=\widehat K_0+\sum_{i=1}^T\widehat\xi_{i}\Delta S_{i}+\widehat L_T\label{FSdKd}
\ee
into a square-integrable $\F_0$-measurable random variable $\widehat K_0$, the terminal value $\sum_{i=1}^T\widehat\xi_{i}\Delta S_{i}$ of a stochastic integral with respect to the price process, and the terminal value of a square-integrable martingale $\widehat L$ which is strongly $P$-orthogonal to $M$. If the integrand $\widehat \xi$ is in $\Theta$ and one replaces the left-hand side by any $H\in L^2(\Om,\F,P)$, a decomposition of the form
$$H=\widehat H_0+\sum_{i=1}^T\widehat\xi^H_{i}\Delta S_{i}+\widehat L^H_T$$
is called the \emph{F\"ollmer--Schweizer (FS) decomposition} of $H$, and the integrand $\widehat\xi^H$ yields the so-called \emph{locally risk minimising strategy} for the contingent claim $H$; see e.g.~\cite{S01} and \cite{S08}. However, it turns out that $\widehat\xi=\lambda-\gamma\hvt$ is in general not in $\Theta$ and therefore \eqref{FSdKd} does not necessarily coincide with the FS decomposition of $K_T$; see Corollary \ref{ces} below for a sufficient condition. But nevertheless, \eqref{FSdKd} gives an \emph{intuitive explanation of the LMVE strategy}. On the one hand, the LMVE investor is optimising  the conditional mean-variance criterion of the gains in the next period only by choosing the \emph{myopically mean-variance efficient strategy (MMVE)}  $\hvp\in\TS$ given by $\widehat\vp_k:=\frac{1}{\gamma}\lambda_k=\frac{1}{\gamma}\frac{E[\Delta S_{k}|\F_{k-1}]}{\Var\left[\Delta S_{k}|\F_{k-1}\right]}$ for $k=1,\ldots,T$. This strategy solves the problem to
\be
\text{\textrm{maximise $U_{k-1}(\vt\mathbbm{1}_{\{k\}})=E[\vt_{k}\Delta S_{k}|\F_{k-1}]-\frac{\gamma}{2}\Var[\vt_{k}\Delta S_{k}|\F_{k-1}]$ over all $\vt\in\TS$}}\label{SMVPS}
\ee
for all $k=1,\ldots,T$. The latter follows immediately as in the one-period case; see \eqref{OPMVPS} and \eqref{opmvest} and also Proposition \ref{prop:MMVE} later. Considering the MMVE strategy in the multiperiod setting the LMVE investor takes by \eqref{dpe} also the fluctuations of the expected future gains into account. The risk resulting from these is due to the stochastic investment opportunity set and can be represented by $\frac{1}{\gamma}K_T$. In addition to holding the MMVE strategy the LMVE investor then minimises the risk resulting from this by local risk minimisation on the other hand which leads to the additional intertemporal hedging demand $\frac{1}{\gamma}\widehat\xi=\widehat\xi(\hvt)$ in the LMVE strategy. As a matter of fact, the intertemporal hedging demand is zero and the LMVE and the MMVE strategy coincide, if the investment opportunity set or more generally the terminal value of the MVT process is deterministic; see Corollary \ref{cordos} below.

Besides this interpretation the above also gives an alternative, in some sense global, characterisation of the LMVE strategy in terms of the structure condition and the MVT process, which is summarised in the next lemma.
 \bl\label{lacdt}
There exists a LMVE strategy $\hvt$ if and only if $S$ satisfies (SC) and (the terminal value of) the MVT process $K_T$ is in $L^1(P)$ and can be written as
\be
K_T=\widehat K_0+\sum_{i=1}^T\widehat\xi_{i}\Delta S_{i}+\widehat L_T\label{FSdd}
\ee
with $\widehat K_0\in L^2(\Om,\cF_0,P)$, $\widehat\xi\in L^2(M)$ such that $\widehat\xi-\lambda\in L^2(A)$, and $\widehat L\in\cM_0^2(P)$ strongly orthogonal to $M$. In that case, $\widehat\vt$ is given by $\widehat\vt=\frac{1}{\gamma}\big(\lambda-\widehat\xi\big)$.\\
If $K_T$ is in $L^2(P)$ and admits a decomposition \eqref{FSdd}, the integrand $\widehat\xi$ is in $\Theta$ and \eqref{FSdd} coincides with the F\"ollmer--Schweizer decomposition of $K_T$.
\el
\bp
By plugging \eqref{doptcond} into \eqref{eq:acdt} and comparing this with \eqref{FSdKd}, we obtain that $\widehat\xi=\lambda-\gamma\hvt=\gamma\xi(\hvt)$ and therefore the first assertion. If $K_T$ is in $L^2(P)$, this gives that $\lambda\in\Theta_S$, which implies that $\widehat\xi\in\TS$ and completes the proof.
\ep
\section{Continuous Time}\label{sec:ct}
In continuous time, we should like to obtain the time-consistent solution in analogy to discrete time by optimising the mean-variance criterion with respect to local perturbations. For a precise formulation of this we need a local description of the underlying quantities and a limit argument. To that end, let us fix some terminology first.

Recall from Section \ref{se:fp} that we can and do assume that $S$ is square-integrable with canonical decomposition $S=S_0+M+A$, where $M$ is an $\R^d$-valued square-integrable martingale null at zero, i.e.~$M\in\cM^2_{0}(P)$, and $A$ is an $\R^d$-valued predictable finite variation RCLL process null at zero. By Propositions II.2.9 and II.2.29 in \cite{JS}, there exist an increasing, integrable, predictable RCLL process $B$, an $\R^d$-valued predictable process $a$ and a predictable $\R^{d\times d}$-valued process $c^M$ whose values are positive semidefinite symmetric matrices such that
\be
\text{$\vt\sint A=(\vt^\T a)\sint B\qquad$ and $\qquad\la \vt\sint M\ra =(\vt^\T c^M \vt) \sint B\qquad$ for all $\vt\in\TS$.}\label{Bac}
\ee
By adding $t$ to $B$, we can assume that $B$ is strictly increasing. Set $P_B:=\PtB$. There exist many processes $B$, $a$ and $c^M$ satisfying \eqref{Bac}, but our results do not depend on the specific choice we make. Using the Moore--Penrose pseudoinverse $(c^M)^{-1}$ of $c^M$ (see \cite{A72}) or the arguments preceding Theorem 2.3 in \cite{DS96}, we define a predictable process $\lambda:=(c^M)^{-1} a$ which gives a decomposition
\be
a=c^M\lambda + \eta\label{da}
\ee
such that $\eta$ is valued in $\Ker(c^M)$. Then $S$ satisfies the \emph{structure condition (SC)} if and only if $\eta=0$ and $\lambda\in L^2_{loc}(M)$, i.e.~the \emph{mean-variance tradeoff (MVT) process} $K$ given by
$K_t=\int_0^t\lambda_u^\T d\la M\ra_u \lambda_u=\la \lambda \sint M\ra_t$ for $t\in[0,T]$ is $P$-a.s.~finite. In continuous time, the process of \emph{expected future gains} $Z(\vt)$ and the square-integrable martingale $Y(\vt)$ of its canonical decomposition are given by \begin{align*}
Z_t(\vt):=E\left[\int_t^T\vt_ud S_u\bigg|\cF_t\right]=E\left[\int_0^T\vt_ud A_u\bigg|\cF_t\right]-\int_0^t\vt_ud A_u=:Y_t(\vt)-\int_0^t\vt_ud A_u
\end{align*}
for $t\in[0,T]$ and each strategy $\vt\in\Theta$. Using the (continuous-time) GKW decomposition
$$\int_0^T\vt_ud A_u=Y_0(\vt)+\int_0^T\xi_u(\vt)d M_u+L_T(\vt)$$
of $Y(\vt)$, we can rewrite $Z(\vt)$ as
\be
Z_t(\vt)=Y_t(\vt)-\int_0^t\vt_ud A_u=Y_0(\vt)+\int_0^t\xi_u(\vt)d M_u+L_t(\vt)-\int_0^t\vt_ud A_u\label{eq:adZct}
\ee
for $t\in[0,T]$, exactly as in discrete time.

A \emph{partition} of $[0,T]$ is a finite set $\tau=\{t_0,t_1,\ldots,t_m\}$ with \mbox{$0=t_0<t_1<\cdots<t_m=T$}, and its \emph{mesh size} is \mbox{$|\tau|:=\max_{t_i\in\tau\setminus\{T\}}(t_{i+1}-t_{i})$.} A sequence of partitions $(\tau_n)_{n\in\N}$ is \emph{increasing} if $\tau_n\subseteq\tau_{n+1}$ for all $n$; it \emph{tends to the identity} if $\lim_{n\to\infty}|\tau_n|=0$. For later use, we associate to each partition $\tau$ the $\sigma$-field
$$\cP^{\tau}:=\sigma\big(\big\{F_0\times\{0\}, F_{i}\times(t_i,t_{i+1}] \big| t_i\in\tau\setminus\{T\}, F_0\in\cF_0, F_{t_i}\in\cF_{t_i} \big\}\big)$$
on $\OmT$. Note for any sequence of partitions $(\tau_n)_{n\in\N}$ tending to the identity that $\sigma\Big(\underset{n\in\N}{\bigcup}\cP^{\tau_n}\Big)$ is equal to the predictable $\sigma$-field $\cP$ and that $\cP^{\tau_n}$ increases to $\cP$ if $(\tau_n)_{n\in\N}$ is in addition increasing. The optimality with respect to local perturbations can then be formulated in continuous time as follows. Recall the notations $U_t(\vt)$ from \eqref{DMVPS} and $\overline{U}_t(\vt)=U_t\big(\mathbbm{1}_{\rrbracket t,T\rrbracket}\vt\big)$.
\begin{defi}
For $\vt,\delta\in\Theta$ and a partition $\tau$ of $[0,T]$, we set
\begin{align}
u^{\tau}[\vt,\delta]&:=\sum_{t_i\in\tau\setminus\{T\}}\frac{U_{t_i}(\vt)-U_{t_i}(\vt+\delta\mathbbm{1}_{(t_i,t_{i+1}]})}{E[B_{t_{i+1}}-B_{t_i}|\F_{t_i}]}\mathbbm{1}_{(t_i,t_{i+1}]}\label{du}\\
&\phantom{:}=\sum_{t_i\in\tau\setminus\{T\}}\frac{\overline{U}_{t_i}(\vt)-\overline{U}_{t_i}(\vt+\delta\mathbbm{1}_{(t_i,t_{i+1}]})}{E[B_{t_{i+1}}-B_{t_i}|\F_{t_i}]}\mathbbm{1}_{(t_i,t_{i+1}]}.\nonumber
\end{align}
A strategy $\hvt\in\Theta$ is called \emph{locally mean-variance efficient (in continuous time)} if
\be
\liminf_{n\to \infty}u^{\tau_n}[\hvt,\delta]\geq 0 \quad P_B\text{-a.e.}\label{lmvect}
\ee
for any increasing sequence $(\tau_n)_{n\in\N}$ of partitions tending to the identity and any $\delta\in\Theta$.
\end{defi}
Intuitively, $u^{\tau}[\vt,\delta]$ measures the change in the tradeoff between mean and variance  of the gains over the remaining time interval  when we perturb $\vt$ locally by $\delta$ along $\tau$. Condition \eqref{lmvect} then says that perturbing the optimal stratetgy $\hvt$ locally should always decrease this tradeoff, at least asymptotically. The appropriate ``time scale'' for this asymptotic is given by the process $B$ which is sometimes also referred to as operational time in the literature. In analogy to discrete time, finding the time-consistent solution by recursive optimisation is captured by comparing at time $t_i$ strategies which differ only on $(t_i,t_{i+1}]$ but are equal on $(t_{i+1},T]$. Passing to the limit then takes this recursive optimisation to continuous time. By the usual embedding of the discrete-time case into the continuous-time setting (as for example explained in Section I.1f in \cite{JS}) it is straightforward to see that the continuous-time formulation \eqref{lmvect} coincides with that in discrete time \eqref{dd2}, since we can choose $B_t=\sum_{k=1}^T\mathbbm{1}_{\{k\leq t\}}$ in this situation (see Section II.3 in \cite{JS}).

The definition of local mean-variance efficiency above as well as the subsequent treatment are inspired by the concept of local risk minimisation in continuous time introduced by Schweizer in \cite{S88}; see also \cite{S01} and \cite{S08}. To obtain a characterisation of the LMVE strategy $\hvt$ we need to derive the asymptotics of \eqref{lmvect}. As in \cite{S08}, the first ingredient for this is a decomposition of $u^{\tau}$ into three terms $A_1^{\tau}$, $A_2^{\tau}$ and $A_3^{\tau}$ for which we can control the asymptotics of each one separately. This follows by using the same arguments as in \cite{S08} which we give here for completeness. 
\begin{prop}\label{pdA}
For all strategies $\vt,\delta\in\Theta$ and every partition $\tau$ of $[0,T]$, we have
$$u^\tau[\vt,\delta]=A^\tau_1+A^\tau_2+A^\tau_3,$$
where
\begin{align*}
A^\tau_1&=E_B\left[\left(\gamma\big(\xi(\vt)+\vt\big)-\lambda-\frac{\gamma}{2}\delta\right)^\T c^M \delta+\delta^\T \eta\Big|\cP^\tau\right]\\
A^\tau_2&=\frac{\gamma}{2}\sum_{t_i\in\tau_n\setminus\{T\}}\frac{\Var\left[\int_{t_i}^{t_{i+1}}\delta dA\Big|\F_{t_i}\right]}{E[B_{t_{i+1}}-B_{t_i}|\F_{t_i}]}\mathbbm{1}_{(t_i,t_{i+1}]}\\
A^\tau_3&=\gamma\sum_{t_i\in\tau_n\setminus\{T\}}\frac{\Cov\left[L_{t_{i+1}}(\vt)-L_{t_i}(\vt)+\int_{t_i}^{t_{i+1}}\big(\xi(\vt)+\vt+\delta\big)dM, \int_{t_i}^{t_{i+1}}\delta dA\Big|\F_{t_i}\right]}{E[B_{t_{i+1}}-B_{t_i}|\F_{t_i}]}\mathbbm{1}_{(t_i,t_{i+1}]}.
\end{align*}
\end{prop}
\bp
Plugging $\vt$ and $\vt+\delta\mathbbm{1}_{(t_i,t_{i+1}]}$ into the definition of $U(\cdot)$ gives that
\begin{align}
&U_{t_i}(\vt)-U_{t_i}(\vt+\delta\mathbbm{1}_{(t_i,t_{i+1}]})\nonumber\\
&=-E\left[\int_{t_i}^{t_{i+1}}\delta_u dS_u\bigg|\F_{t_i}\right]+\gamma\Cov\left[\int_{0}^T\vt_u dS_u+\frac{1}{2}\int_{t_i}^{t_{i+1}}\delta_u dS_u,\int_{t_i}^{t_{i+1}}\delta_u dS_u\Bigg|\F_{t_i}\right].\label{p:eq1}
\end{align}
Using $S=S_0+M+A$ and the definition of $Y(\vt)$ we can write
$$\int_{0}^T\vt_u dS_u-E\left[\int_{0}^T\vt_u dS_u\bigg|\F_{t_i}\right]=Y_T(\vt)-Y_{t_i}(\vt)+\int_{t_i}^T\vt_udM_u,$$
which gives
\begin{align}
&\Cov\left[\int_{0}^T\vt_u dS_u+\frac{1}{2}\int_{t_i}^{t_{i+1}}\delta_u dS_u\, ,\int_{t_i}^{t_{i+1}}\delta_u dS_u\Bigg|\F_{t_i}\right]\nonumber\\
&=\Cov\left[Y_T(\vt)-Y_{t_i}(\vt)+\int_{t_i}^{t_{i+1}}\left(\vt_u+\frac{1}{2}\delta_u\right) dM_u,\int_{t_i}^{t_{i+1}}\delta_u dM_u\Bigg|\F_{t_i}\right]\nonumber\\
&\phantom{=}+\Cov\left[Y_T(\vt)-Y_{t_i}(\vt)+\int_{t_i}^{t_{i+1}}\left(\vt_u+\delta_u\right) dM_u,\int_{t_i}^{t_{i+1}}\delta_u dA_u\Bigg|\F_{t_i}\right]\nonumber\\
&\phantom{=}+\frac{1}{2}\Var\left[\int_{t_i}^{t_{i+1}}\delta_u dA_u\bigg|\F_{t_i}\right].\label{p:eq2}
\end{align}
Since $Y(\vt)$ and $\int \vt dM$ are martingales, the second term on the right-hand side above equals
\be
\Cov\left[Y_{t_{i+1}}(\vt)-Y_{t_i}(\vt)+\int_{t_i}^{t_{i+1}}\left(\vt_u+\delta_u\right) dM_u,\int_{t_i}^{t_{i+1}}\delta_u dA_u\Bigg|\F_{t_i}\right].\label{p:eq3}
\ee
With an analogous argument and inserting the Galtchouk--Kunita--Watanabe decomposition $Y(\vt)=Y_0(\vt)+\int\xi(\vt)dM+L(\vt),$ we obtain
\begin{align}
&\Cov\left[Y_T(\vt)-Y_{t_i}(\vt)+\int_{t_i}^T\vt_udM_u+\frac{1}{2}\int_{t_i}^{t_{i+1}}\delta_u dM_u,\int_{t_i}^{t_{i+1}}\delta_u dM_u\Bigg|\F_{t_i}\right]\nonumber\\
&=\Cov\left[\int_{t_i}^{t_{i+1}}(\xi_u(\vt)+\vt_u)dM_u+L_{{t_{i+1}}}(\vt)-L_{t_i}(\vt)+\frac{1}{2}\int_{t_i}^{t_{i+1}}\delta_u dM_u,\int_{t_i}^{t_{i+1}}\delta_u dM_u\Bigg|\F_{t_i}\right]\nonumber\\
&=E\left[\int_{t_i}^{t_{i+1}}d\Big\la \mbox{$\int \left(\xi(\vt)+\vt+\frac{1}{2}\delta\right)dM$},\mbox{$\int\delta dM$}\Big\ra \Bigg|\F_{t_i}\right]\nonumber\\
&=E\left[\int_{t_i}^{t_{i+1}}\left(\mbox{$\xi(\vt)_u+\vt_u+\frac{1}{2}\delta_u$}\right)^\T c^M_u \delta_udB_u \Bigg|\F_{t_i}\right].\label{p:eq4}
\end{align}
By the martingale property of $\int \delta dM$ and using $a=c^M\lambda+\eta$ we have
\be
E\left[\int_{t_i}^{t_{i+1}}\delta_u dS_u\bigg|\F_{t_i}\right]=E\left[\int_{t_i}^{t_{i+1}}(\delta _u^\T c^M_u\lambda_u + \delta_u^\T \eta_u )dB_u\bigg|\F_{t_i}\right].\label{p:eq5}
\ee
Combining \eqref{p:eq1}--\eqref{p:eq5} we conclude that 
\begin{align*}
&U_{t_i}(\vt)-U_{t_i}(\vt+\delta|_{(t_i,t_{i+1}]})\\
&= E\left[\int_{t_i}^{t_{i+1}}\left(\left(\mbox{$\gamma\big(\xi(\vt)_u+\vt_u\big)-\lambda_u+\frac{\gamma}{2}\delta_u$}\right)^\T c^M_u \delta_u-\delta_u^\T \eta_u\right)dB_u \Bigg|\F_{t_i}\right]\\
&\phantom{=}+\gamma\Cov\left[Y_{t_{i+1}}(\vt)-Y_{t_i}(\vt)+\int_{t_i}^{t_{i+1}}\left(\vt_u+\delta_u\right) dM_u,\int_{t_i}^{t_{i+1}}\delta_u dA_u\Bigg|\F_{t_i}\right]\\
&\phantom{=}+\frac{\gamma}{2}\Var\left[\int_{t_i}^{t_{i+1}}\delta_u dA_u\bigg|\F_{t_i}\right].
\end{align*}
After dividing by $E[B_{t_{i+1}}-B_{t_{i}}|\cF_{t_i}]$, multiplying by $\mathbbm{1}_{(t_i,t_{i+1}]}$ and summing over $t_i\in\tau\setminus\{T\}$, we obtain $u^\tau[\vt,\delta]$ on the left-hand side and $A^\tau_1$, $A^\tau_3$ and $A^\tau_2$ on the right-hand side, as
\begin{align*}
&\sum_{t_i\in\tau_n\setminus\{T\}}\frac{E\left[\int_{t_i}^{t_{i+1}}\left(\left(\mbox{$\gamma\big(\xi(\vt)_u+\vt_u\big)-\lambda_u+\frac{\gamma}{2}\delta_u$}\right)^\T c^M_u \delta_u-\delta_u^\T \eta_u\right)dB_u \Big|\F_{t_i}\right]}{E[B_{t_{i+1}}-B_{t_{i}}|\cF_{t_i}]}\mathbbm{1}_{(t_i,t_{i+1}]}\\
&=E_B\left[\left(\mbox{$\xi(\vt)+\vt-\lambda+\frac{1}{2}\delta$}\right)^\T c^M \delta+\delta^\T \eta\Big|\cP^\tau\right]=A^\tau_1, 
\end{align*}
which completes the proof.
\ep
Since $A_1^\tau$ is of the same form as the corresponding term in Proposition 2.2 in \cite{S08}, we obtain its asymptotic behaviour by the same argument as in Lemma 3.1 in \cite{S08}. The additional term $\delta^\T\eta$ is not relevant for this.
\bl\label{lA_1}
Let $(\tau_n)_{n\in\N}$ be an increasing sequence of partitions tending to the identity. Then
\be
\lim_{n\to\infty}A^{\tau_n}_1=\left(\gamma\big(\xi(\vt)+\vt\big)-\lambda+\frac{\gamma}{2}\delta\right)^\T c^M \delta-\delta^\T \eta\qquad P_B\text{-a.e.}\label{l:eq1}
\ee
\el
\bp
We observe that $\left(\mbox{$\gamma\big(\xi(\vt)+\vt\big)-\lambda+\frac{1}{2}\delta$}\right)^\T c^M \delta-\delta^\T \eta\in L^1(P_B)$, since $\vt$ and $\delta$ are in $\Theta$, and recall that $(\cP^{\tau_n})_{n\in\N}$ increases to the predictable $\sigma$-field $\cP$, since $(\tau_n)_{n\in\N}$ is increasing and tending to the identity. As $A^{\tau_n}_1=E_B\big[\big(\mbox{$\gamma\big(\xi(\vt)+\vt\big)-\lambda+\frac{1}{2}\delta$}\big)^\T c^M \delta-\delta^\T \eta\big|\cP^{\tau_n}\big]$ by definition, $(A^{\tau_n}_1)_{n\in\N}$ is a uniformly integrable $P_B$-martingale and \eqref{l:eq1} follows from the martingale convergence theorem, since $\left(\mbox{$\gamma\big(\xi(\vt)+\vt\big)-\lambda+\frac{1}{2}\delta$}\right)^\T c^M \delta-\delta^\T \eta$ is predictable.
\ep
To show that the term $A^{\tau_n}_2$ is asymptotically negligible, we establish the following general convergence result. For this we argue with the predictable measurability of $X$ and need not assume continuity of $X$ as in Proposition 3.5 in \cite{S08}. Applying our techniques to \emph{local risk minimisation} enables us to generalise this concept and some related results to a general semimartingale setting as well. In particular, we are able to drop the continuity of $A$ and (SC) in Theorem 1.6 and Proposition 5.2 in \cite{S08}; this will be explained in more detail in future work.
\bl\label{lA_3}
Let $(\tau_n)_{n\in\N}$ be an increasing sequence of partitions of $[0,T]$ tending to the identity and $X\in\cH^2(P)$  a predictable finite variation process such that $X=\int \alpha dB$ for $\alpha\in L^0(B)$. Then
\be
\lim_{n\to\infty}\sum_{t_i\in\tau_n\setminus\{0\}}\frac{\Var\left[X_{t_i}-X_{t_{i-1}}|\F_{t_{i-1}}\right]}{E[B_{t_{i}}-B_{t_{i-1}}|\F_{t_{i-1}}]}\mathbbm{1}_{(t_{i-1},t_{i}]}=0\qquad P_B\text{-a.e.}\label{l2:eq1}
\ee
\el
\bp
We first decompose
\begin{align*}
&\sum_{t_i\in\tau_n\setminus\{0\}}\frac{\Var\left[X_{t_i}-X_{t_{i-1}}|\F_{t_{i-1}}\right]}{E[B_{t_{i}}-B_{t_{i-1}}|\F_{t_{i-1}}]}\mathbbm{1}_{(t_{i-1},t_{i}]}\\
&=\sum_{t_i\in\tau_n\setminus\{0\}}\frac{E\left[(X_{t_i}-X_{t_{i-1}})^2|\F_{t_{i-1}}\right]}{E[B_{t_{i}}-B_{t_{i-1}}|\F_{t_{i-1}}]}\mathbbm{1}_{(t_{i-1},t_{i}]}-\sum_{t_i\in\tau_n\setminus\{0\}}\frac{\left(E[X_{t_i}-X_{t_{i-1}}|\F_{t_{i-1}}]\right)^2}{E[B_{t_{i}}-B_{t_{i-1}}|\F_{t_{i-1}}]}\mathbbm{1}_{(t_{i-1},t_{i}]}.
\end{align*}
For the proof of \eqref{l2:eq1} we then only need to show that both sums on the right-hand side converge to the same limit $\alpha\Delta X $. To that end, set $t^{\tau_n}=\inf\{s\in\tau_n~|~s\geq t\}$ and $t^{\tau_n-}=\sup\{s\in\tau_n~|~s< t\}$ for each $t\in[0,T]$, and $X^n\omt=(X_{t^{\tau_n}}-X_{t^{\tau_n-}})(\om)$ and $\widetilde X^n\omt=E[X^n_t|\cF_{t^{\tau_n-}}](\om)$ for all $\omt\in\OmT$. Using $X=\int\alpha dB$ we can write
\begin{align*}
&\sum_{t_i\in\tau_n\setminus\{0\}}\frac{E[(X_{t_i}-X_{t_{i-1}})^2|\F_{t_{i-1}}]}{E[B_{t_{i}}-B_{t_{i-1}}|\F_{t_{i-1}}]}\mathbbm{1}_{(t_{i-1},t_{i}]}\\
&=\sum_{t_i\in\tau_n\setminus\{0\}}\frac{E[(X_{t_i}-X_{t_{i-1}})\int_{t_{i-1}}^{t_{i}}\alpha_u dB_u|\F_{t_{i-1}}]}{E[B_{t_{i}}-B_{t_{i-1}}|\F_{t_{i-1}}]}\mathbbm{1}_{(t_{i-1},t_{i}]}=E_B[X^n\alpha|\cP^{\tau_n}]
\end{align*}
and
\begin{align*}
&\sum_{t_i\in\tau_n\setminus\{0\}}\frac{\left(E[X_{t_i}-X_{t_{i-1}}|\F_{t_{i-1}}]\right)^2}{E[B_{t_{i}}-B_{t_{i-1}}|\F_{t_{i-1}}]}\mathbbm{1}_{(t_{i-1},t_{i}]}\\
&=\sum_{t_i\in\tau_n\setminus\{0\}}E[X_{t_i}-X_{t_{i-1}}|\F_{t_{i-1}}]\frac{E\left[\int_{t_{i-1}}^{t_{i}}\alpha_u dB_u\Big|\F_{t_{i-1}}\right]}{E[B_{t_{i}}-B_{t_{i-1}}|\F_{t_{i-1}}]}\mathbbm{1}_{(t_{i-1},t_{i}]}=\widetilde X^nE_B[\alpha|\cP^{\tau_n}].\label{p2:eq1}
\end{align*}
By estimating $\sup_{n\in\N}|X^n\alpha|\leq 2|\alpha|\sup_{0\leq s\leq T}|X_s|$ and $\sup_{0\leq s\leq T}|X_s|\leq\int_0^T|dX_u|$, we obtain that $\sup_{n\in\N}|X^n\alpha|\in L^1(P_B)$ as $\int_0^T(\int_0^T|dX_s|)|\alpha_u|dB_u=\big(\int_0^T|dX_s|\big)^2\in L^1(P)$. Since $X$ is RCLL and $t^{\tau_n}\searrow t$ and $t^{\tau_n-}\nearrow t$ as $n\to\infty$, it follows that $X^n$ converges pointwise to $\Delta X$. Combining this with the integrability of $\sup_{n\in\N}|X^n\alpha|$ gives that $E_B[X^n\alpha|\cP^{\tau_n}]$ tends to $\alpha\Delta  X$ $P_B$-a.e.~by Hunt's lemma (see \cite{DM82}, V.45), since $\cP^{\tau_n}$ increases to $\cP$ and $\alpha\Delta  X$ is predictable. As the $P_B$-a.e.~convergence of $E_B[\alpha|\cP^{\tau_n}]$ to $\alpha$ already follows by the martingale convergence theorem, it remains to show that $\widetilde X^n$ converges to $\Delta X$  $P_B$-a.e.~for the convergence of the second sum. Since $\sup_{n\in\N}|X_{t^{\tau_n}}-X_{t^{\tau_n}-}|\leq 2\int_0^T|dX_s|\in L^2(P)$ for all $t\in[0,T]$ and $X^n$ converges pointwise to $\Delta X$, it follows by Hunt's lemma that
\be
\widetilde X^n_t \longrightarrow E[\Delta X_t|\cF_{t-}]\quad P\text{-a.s.~for each $t\in[0,T]$.}\label{p2:eq2}
\ee
By Theorem III.5 in \cite{P04} the limit coincides with $\Delta X_t$, as $\Delta X$ is predictable. Since $\{\lim_{n\to\infty}\widetilde X^n\ne \Delta X \}\in\cF\otimes\mathcal{B}([0,T])$, we obtain that $\widetilde X^n$ converges to $\Delta X$ $P_B$-a.e.~from \eqref{p2:eq2} by Fubini's theorem. This completes the proof.
\ep
With this we have now everything in place to derive the asymptotics of $u^\tau[\vt,\delta]$.
\bl\label{lconvu}
Let $(\tau_n)_{n\in\N}$ be an increasing sequence of partitions of $[0,T]$ tending to the identity. Then
$$\lim_{n\to\infty}u^{\tau_n}[\vt,\delta]=\left(\gamma\big(\xi(\vt)+\vt\big)-\lambda+\frac{\gamma}{2}\delta\right)^\T c^M \delta-\delta^\T \eta\qquad P_B\text{-a.e.}
$$
for all $\vt,\delta\in\Theta$.
\el
\bp
The proof follows immediately by combining Proposition \ref{pdA} and Lemma \ref{lA_1} after we have shown that $A^{\tau_n}_2$ and $A^{\tau_n}_3$ converge to $0$ $P_B$-a.e. To that end, we estimate
\begin{align*}
&\left|\Cov\left[Y_{t_{i+1}}(\vt)-Y_{t_i}(\vt)+\int_{t_i}^{t_{i+1}}\left(\vt_u+\delta_u\right) dM_u,\int_{t_i}^{t_{i+1}}\delta_u dA_u\bigg|\F_{t_i}\right]\right|^2\\
&\leq \Var\left[Y_{t_{i+1}}(\vt)-Y_{t_i}(\vt)+\int_{t_i}^{t_{i+1}}\left(\vt_u+\delta_u\right) dM_u\bigg|\F_{t_i}\right]\Var\left[\int_{t_i}^{t_{i+1}}\delta_u dA_u\bigg|\F_{t_i}\right]\\
&=E\big[X_{t_{i+1}}-X_{t_{i+1}}\big|\F_{t_i}\big]\Var\left[\int_{t_i}^{t_{i+1}}\delta_u dA_u\bigg|\F_{t_i}\right]
\end{align*}
by using the Cauchy-Schwarz inequality and $X:=\big\la Y+\int (\vt+\delta)dM \big\ra$. Again by the Cauchy-Schwarz inequality we obtain from the above that
\begin{eqnarray}
|A_3^{\tau_n}|&\leq &\gamma\left(\sum_{t_i\in\tau_n\setminus\{T\}}\frac{E\big[X_{t_{i+1}}-X_{t_{i+1}}\big|\F_{t_i}\big]}{E[B_{t_{i+1}}-B_{t_i}|\F_{t_i}]}\mathbbm{1}_{(t_i,t_{i+1}]}\right)^{\frac{1}{2}}\nonumber\\
&&\times\left(\sum_{t_i\in\tau_n\setminus\{T\}}\frac{\Var\left[\int_{t_i}^{t_{i+1}}\delta_udA_u\Big|\F_{t_i}\right]}{E[B_{t_{i+1}}-B_{t_i}|\F_{t_i}]}\mathbbm{1}_{(t_i,t_{i+1}]}\right)^{\frac{1}{2}}\nonumber\\
&=&\sqrt{2\gamma}\left(\frac{dP_X}{dP_B}\Big|_{\cP^{\tau_n}}\right)^{\frac{1}{2}}(A^{\tau_n}_2)^{\frac{1}{2}},\label{p:lconvu:eq1}
\end{eqnarray}
where $P_X:=P\otimes X$ and $\frac{dP_X}{dP_B}\big|_{\cP^{\tau_n}}=\sum_{t_i\in\tau_n\setminus\{T\}}\frac{E[X_{t_{i+1}}-X_{t_{i}}|\F_{t_i}]}{E[B_{t_{i+1}}-B_{t_i}|\F_{t_i}]}\mathbbm{1}_{(t_i,t_{i+1}]}$. It is straightforward to verify that $\big(\frac{dP_X}{dP_B}\big|_{\cP^{\tau_n}}\big)_{n\in\N}$ is a $P_B$-martingale by simply checking the definition; see Lemma 3.4 in \cite{S08}. Since $\frac{dP_X}{dP_B}\big|_{\cP^{\tau_n}}$ is non-negative, it follows directly by the martingale convergence theorem that $\big(\frac{dP_X}{dP_B}\big|_{\cP^{\tau_n}}\big)_{n\in\N}$ is $P_B$-a.e.~convergent and hence $P_B$-a.e.~bounded in $n$. (Moreover, the limit coincides with the Radon--Nikod\'ym derivative of the absolutely continuous part of $P_X$ with respect to $P_B$.) Since $\int\delta dA=\int \delta^\T adB$, applying Lemma \ref{lA_3} with $\alpha=\delta^\T a$ yields that $\lim_{n\to\infty}A_2^{\tau_n}=0$ $P_B$-a.e. and therefore also that $\lim_{n\to\infty}A_3^{\tau_n}=0$ $P_B$-a.e.~by \eqref{p:lconvu:eq1}. This completes the proof.
\ep
Having the representation of our criterion above, we can now describe the solution.
\bt\label{thmct}

There exists a LMVE strategy $\hvt$  if and only if we have both
\bi
\item[1)] $S$ satisfies (SC) with $\lambda\in L^2(M)$, i.e.~$K_T\in L^1(P)$.
\item[2)] There exists $\widehat\psi\in\Theta$ such that
\be
\widehat\psi=\frac{1}{\gamma}\lambda -\xi(\widehat\psi)\label{optcond},
\ee
where $\xi(\widehat\psi)$ is the integrand in the GKW decomposition of $\int_0^T\widehat\psi_udA_u$.
\ei
In that case, $\hvt=\widehat\psi$.
\et
\bp
Using Lemma \ref{lconvu} it follows by definition that $\hvt$ is LMVE if and only if 
\be
\left(\gamma\big(\xi(\hvt)+\hvt\big)-\lambda+\frac{\gamma}{2}\delta\right)^\T c^M \delta-\delta^\T \eta\geq 0\qquad P_B\text{-a.e.}\label{pr:optcond:ct}
\ee
for all $\delta\in\Theta$. If 1) and 2) hold, \eqref{pr:optcond:ct} reduces to $\frac{\gamma}{2}\delta^\T c^M \delta\geq 0$ for $\hvt:=\widehat\psi=\frac{1}{\gamma}\lambda- \xi(\widehat\psi)$ and all $\delta\in\TS$, which immediately gives that this strategy $\hvt$ is LMVE. For the converse, we first observe that since $c^M \eta=0$, choosing $\delta=\eta\mathbbm{1}_{\{|\eta^\T a| \leq n\}}$ for each $n\in\N$ gives that $\delta\in\TS$ and $-\delta^\T\delta\geq 0$ in \eqref{pr:optcond:ct}. This implies that $\eta=0$ $P_B$-a.e.~and therefore that $S$ satisfies (SC). Set $\vp=\frac{1}{\gamma}\lambda-\big(\xi(\hvt)+\hvt\big)$. Then plugging $\delta=\vp\mathbbm{1}_{\{\vp^\T c^M\vp +|\vp^\T a| \leq n\}}\in\TS$ into \eqref{pr:optcond:ct} for each $n\in\N$ yields that $-\frac{\gamma}{2}\vp^\T c^M\vp\geq 0$ $P_B$-a.e.~so that $\vp=0$ in $L^2(M)$, which gives that $\lambda=\gamma\big(\xi(\hvt)+\vt\big)\in L^2(M)$. This completes the proof.
\ep
As in discrete time, we say that a random variable $H\in L^2(\Om,\F_T,P)$ admits a \emph{F\"ollmer--Schweizer decomposition} if it can be written as
\be
H=\widehat H_0+\int_0^T\widehat\xi^H_ud S_{u}+\widehat L^H_T,\label{defFSct}
\ee
where $\widehat H_0\in L^2(\Om,\cF_0,P)$, $\widehat \xi^H\in\Theta$ and $\widehat L^H\in\cM^2_0(P)$ is strongly $P$-orthogonal to $M$. However, unlike the discrete-time case a FS decomposition in continuous time is no longer unique in general; see Remark 1.4 in \cite{CS96} and Example \ref{ex:non-uniqueness} below. Using this notion we can then give the following alternative characterisation of the LMVE. Note that in contrast to the notion of optimality this alternative description is in some sense global. 
\bt\label{thmLMVEFS}
There exists a LMVE strategy $\hvt$ if and only if $S$ satisfies (SC) and (the terminal value of)  the MVT process $K_T$ is in $L^1(P)$ and can be written as
\be
K_T=\widehat K_0+\int_0^T\widehat\xi d S+\widehat L_T\label{FSdc}
\ee
with $\widehat K_0\in L^2(\Om,\cF_0,P)$, $\widehat\xi\in L^2(M)$ such that $\widehat\xi-\lambda\in L^2(A)$, and $\widehat L\in\cM_0^2(P)$ strongly $P$-orthogonal to $M$. In that case, $\widehat\vt$ is given by $\widehat\vt=\frac{1}{\gamma}\big(\lambda-\widehat\xi\big)$, $\xi(\hvt)=\frac{1}{\gamma}\widehat\xi$,
\be
Z_t(\hvt)=\frac{1}{\gamma}\left(\widehat K_0+\int_0^t\widehat\xi dS+\widehat L_t-K_t\right)\label{dZct}
\ee
and
\begin{align}
U_t(\widehat \vt)&=x+\int_0^t\left(\widehat \vt+\frac{1}{\gamma}\widehat\xi\right)dS+\frac{1}{\gamma}\left(\widehat K_0+\widehat L_t-\frac{1}{2}E\left[K_T-K_t+\big\la \widehat L\big\ra_T-\big\la \widehat L\big\ra_t\Big|\F_t\right]\right)\label{dUct}
\end{align}
with canonical decomposition
\be
U_t(\widehat \vt)=x+\frac{1}{\gamma}\left(\widehat K_0+\int_0^t\lambda dM+\widehat L_t-\frac{1}{2}E\left[K_T+\big\la \widehat L\big\ra_T\Big|\F_t\right]\right)+\frac{1}{2\gamma}\left(K_t+\big\la\widehat L\big\ra_t\right).\label{cdUct}
\ee
If $K_T$ is in $L^2(P)$ and admits a decomposition \eqref{FSdc}, the integrand $\widehat\xi$ is in $\Theta$ and \eqref{FSdc} coincides with the F\"ollmer--Schweizer decomposition of $K_T$.
\et
\bp
The equivalence between the existence of the LMVE strategy $\hvt$ and the decomposition \eqref{FSdc} follows from Theorem \ref{thmct} by the same arguments as in discrete time given in the proof of Lemma \ref{lacdt} and before. Indeed by comparing \eqref{eq:acdt} and \eqref{FSdKd}, the integrability properties can be ticked off from the corresponding parts in the decomposition, since $K_T=\int_0^T\lambda_u^\T d\la M \ra_u\lambda_u$ is in $L^1(P)$ or $L^2(P)$, respectively. This also yields \eqref{dZct} by simply plugging $\hvt=\frac{1}{\gamma}(\lambda-\widehat\xi)$ and the parts of \eqref{FSdc} into \eqref{eq:adZct}. For the proof of \eqref{cdUct}, we observe that the square-integrable martingale $R(\hvt)$ given by $R_t(\hvt)=E\big[\int_{0}^T\hvt_u dS_u\big|\F_{t}\big]$ for $t\in[0,T]$ is equal to $\frac{1}{\gamma}(\widehat K_0+\lambda\sint M+\widehat L)$. Inserting this into the definition of $U_t(\hvt)$ gives
\begin{align*}
U_t(\hvt)&=x+R_t(\hvt)-\frac{\gamma}{2}E\Big[\big(R_T(\hvt)-R_t(\hvt)\big)^2\Big|\cF_t\Big]\\
&=x+R_t(\hvt)-\frac{\gamma}{2}E\left[\big\la R(\hvt)\big\ra_T-\big\la R(\hvt) \big\ra_t\Big|\cF_t\right]\\
&=x+\frac{1}{\gamma}(\widehat K_0+\lambda\sint M_t+\widehat L_t)-\frac{1}{2\gamma}E\Big[\la \lambda\sint M\ra_T-\la \lambda\sint M\ra_t+\la \widehat L\ra_T-\la\widehat L \ra_t\Big|\cF_t\Big]
\end{align*}
and therefore \eqref{cdUct}. Since $R_t(\hvt)=\int_{0}^t\hvt_u dS_u+\frac{1}{\gamma}\left(\widehat K_0+\int_0^t\widehat\xi_udS_u+\widehat L_t-K_t\right)$ by \eqref{dZct}, we then obtain \eqref{dUct} from \eqref{cdUct}, which completes the proof.
\ep
In specific Markovian frameworks, relations like in Theorem \ref{thmLMVEFS} have been obtained in \cite{BC} and \cite{BM} by arguments using the Feynman-Kac formula, which are available there. The link between the LMVE strategy $\hvt$ and the FS decomposition now allows us to exploit known results on the FS decomposition to give a sufficient condition for the existence  and uniqueness  of $\hvt$  as well as an example where it is not unique below.  To formulate this, we first need to introduce some of the terminology used in \cite{CKS98}. Since the existence of $\hvt$ implies that $S$ satisfies (SC) with $\lambda\in L^2(M)$, we have that $-\lambda\sint M$ is a square-integrable martingale. For any stopping time $\sigma$ we denote $^\sigma\E(-\lambda\sint M)=\E\big(-(\lambda\mathbbm{1}_{]\mskip-2mu]\sigma,T]\mskip-2mu]})\sint M\big)$. Since $-\lambda\sint M$ is RCLL, it has $P$-a.s.~at most a countable number of jumps with $\Delta(-\lambda\sint M)=-1$, and so we can define an increasing sequence of stopping times $\hat T_n$ by $\hat T_0=0$ and $\hat T_{n+1}=\inf\{t>\hat T_n~|~^{\hat T_n}\E(-\lambda\sint M)_t=0\}\wedge T.$
\begin{defi}
We call $\E(-\lambda\sint M)$ \emph{regular} if for any $n$, $^{\hat T_n}\E(-\lambda\sint M)$ is a martingale.
\end{defi}
\begin{defi} We say that $\E(-\lambda\sint M)$ satisfies \emph{the reverse H\"older inequality} $R_2(P)$, if there exists a constant $c\geq1$ such that for any $t$,
$$E\left[|^t\E(-\lambda\sint M)_T|^2 \big|\F_t\right]\leq c.$$
\end{defi}
\begin{defi}\label{defEmart}
We say that an RCLL process $X$ is an \emph{$\E(-\lambda\sint M)$-martingale}, if for any $n\in\N$,
$$E\big[|X_{\hat T_n}\ ^{\hat T_n}\E(-\lambda\sint M)_{\hat T_{n+1}}|\big]<+\infty$$
and $(\mathbbm{1}_{\rrbracket \hat T_n,T\rrbracket}\sint X)\ ^{\hat T_n}\E(-\lambda\sint M)$ is a martingale.
\end{defi}
\begin{defi} A local martingale $N\in\cM^2_{loc}(P)$ is in \emph{$bmo_2$}, if there exists a constant $c$
such that 
$$E\left[\la N\ra_T -\la N\ra_t|\F_t\right]\leq c^2$$
for all $t\in[0,T]$. The smallest such constant $c$ is denoted by $\|N\|_{bmo_2}$.
\end{defi}
With the definitions above we can give the following sufficient condition for the existence of the LMVE strategy.
\begin{cor}\label{ces}
Suppose that $S$ satisfies (SC) and that $\E(-\lambda\sint M)$ is regular and satisfies $R_2(P)$. Then the LMVE strategy $\hvt$ exists, is  unique and  given by $\hvt=\frac{1}{\gamma}\big(\lambda-\widehat\xi\big)$, where $\widehat\xi\in\TS$ is the  unique  integrand in the FS decomposition of $K_T\in\LiiP$, and
\be
Z_t(\hvt)=\frac{1}{\gamma}E\big[\E\big(-(\lambda\mathbbm{1}_{\rrbracket t,T\rrbracket})\sint M\big)_T(K_T-K_t)\big|\cF_t\big]\label{dZMMM}
\ee
for $t\in[0,T]$. 
\end{cor}
\bp
By Proposition 3.10 in \cite{CKS98}, we have that $-\lambda\sint M$ is in $bmo_2$ and therefore that $K_T=\la \lambda\sint M\ra_T$ is in $L^2(P)$ because $\E(-\lambda\sint M)$ is regular and satisfies $R_2(P)$. Moreover, by Theorem 5.5 in \cite{CKS98}, $S$ admits an FS decomposition (in the stronger sense of Definition 5.4 in \cite{CKS98}), which implies in particular that every $H\in L^2(P)$ has a  unique  FS decomposition, if and only if $\E(-\lambda\sint M)$ is regular and satisfies $R_2(P)$. Combining this with Theorem \ref{thmLMVEFS} we obtain that the LMVE strategy $\hvt$ exists and can be represented as above in terms of the FS decomposition of $K_T$. Since a random variable admits an FS decomposition if and only if it is the terminal value of an $\E$-martingale in $\cH^2(P,\FF)$ (see the discussion preceding Theorem 5.5 in \cite{CKS98}), we obtain that
$$
E\big[\E\big(-(\lambda\mathbbm{1}_{\rrbracket t,T\rrbracket})\sint M\big)_TK_T\big|\cF_t\big]=\widehat K_0+\int_0^t\widehat\xi_udS_u+\widehat L_t
$$
by Proposition 3.12.i) in \cite{CKS98} and therefore \eqref{dZMMM} via \eqref{dZct}, which completes the proof.
\ep
\begin{remark}
\bi
\item[\textbf{1)}] If $\E(-\lambda\sint M)$ is strictly positive in addition to the assumptions above, then it is the density process of an equivalent martingale measure for $S$, the so-called \emph{minimal martingale measure (MMM)} $\widehat P$; see \cite{SMMM}. In this case, \eqref{dZMMM} can be written as $Z_t(\hvt)=\frac{1}{\gamma}\widehat E[K_T-K_t|\cF_t]$. This relation has been obtained in \cite{BC} and \cite{BM} in the specific Markovian frameworks used there by arguments using the Feynman-Kac formula. \item[\textbf{2)}] If the MMM exists and its density process satisfies $R_2(P)$ and $S$ is continuous, then the FS decomposition coincides with the GKW decomposition under $\widehat P$; see \cite{CS96}. In the case, where $S$ is discontinuous, the relation between the two decompositions is more complicated and has recently been established in \cite{CVV10}.
\item[\textbf{3)}] Applying the previous results allows us to obtain the LMVE strategy in concrete models in the following way. First, we check if $S$ satisfies (SC) by using its canonical decomposition. If this is true, we obtain $\lambda$ and therefore $K$ and $\E(-\lambda\sint M)$ directly and explicitly from the canonical decomposition of $S$. If $\E(-\lambda\sint M)$ is regular and satisfies $R_2(P)$, we can try to obtain the FS decomposition of $K_T$ via Theorem 4.3 in \cite{CVV10}, which gives the LMVE strategy by Theorem \ref{thmLMVEFS}. Moreover, if $\E(-\lambda\sint M)$, the candidate for the density process of the MMM, is strictly positive in addition to the previous assumptions, the MMM exists and we can derive the FS decomposition as explained in the previous remark from the GKW decomposition of $K_T$ under $\widehat P$.  In the case that $\E(-\lambda\sint M)$ does not satisfy $R_2(P)$ and is (as in Example \ref{ex:non-uniqueness} below) not even regular, the procedure above can still be used to derive a candidate for the LMVE strategy that might be verified directly to be square-integrable and hence to yield the optimal strategy. 
\item[\textbf{4)}] Since one can obtain the ingredients $\lambda$, $K$ and $\E(-\lambda\sint M)$ directly and explicitly from the canonical decomposition of $S$, obtaining (a candidate for) the LMVE strategy as explained in 3) is more explicit than solving the static but multiperiod or continuous-time Markowitz problem via finding the \emph{variance-optimal martingale measure}; see \cite{S08} and Section 1.3 and 2 of \cite{BC} for a comparison of both strategies in a complete market and a discussion. 

\item[\textbf{5)}] If one cannot determine the LMVE strategy along the steps in part 3) explicitly, one can still try to compute it numerically. For this, one observes that the FS decomposition \eqref{defFSct} is the solution to a linear backward stochastic differential equation (BSDE). If the required conditions are satisfied, one can apply the numerical schemes that have been developed for Markovian Lipschitz BSDEs to solve the BSDE numerically; see \cite{BET09} for an overview as well as the references therein. As the BSDE is linear, these algorithms simplify to calculating conditional expectations and integrands in martingale representations numerically. This has already been observed in \cite{BC}, where it has been suggested to do this with Monte Carlo simulation and Malliavin derivatives.  
\ei 
\end{remark}

The following example illustrates how one can calculate a LMVE strategy explicitly and shows that it might not be unique in general. The example uses the same idea as that for the non-uniqueness of the FS decomposition in \cite{CS96}.
\begin{example}\label{ex:non-uniqueness}
There exists a price process $S\in\cH^2(P)$ such that $S$ satisfies (SC), the terminal value of the MVT process admits a FS decomposition and hence the LMVE strategy exists. However, the integrand in the FS decomposition and the LMVE strategy are not unique. Moreover, the solution to the static MVPS problem \eqref{MVPS} fails to exist, as the price process does not admit an (equivalent) martingale measure.

For convenience we give the construction on the infinite time interval $[0,+\infty)$. The corresponding example on the finite interval $[0,T]$ can be easily obtained from that by using the time change $h:[0,+\infty)\to [0,T)$ given by $h(t)=T\big(1-\exp(-t)\big)$ and then considering $S_{h(t)}$ instead of $S_t$. 

Let $W=(W_t)_{t\geq 0}$ be a Brownian motion on $[0,+\infty)$ and set
$$\textstyle\sigma:=\inf\{t>0~|~\E(-W)_t=\frac{1}{2}\}=\inf\{t>0~|~W_t+\frac{1}{2}t=\log 2\}$$
and $S_t:=W_{\sigma\wedge t}+\sigma\wedge t$. Since $[M^S]_\infty=\sigma$ and $\int_0^\infty|dA_u^S|=\sigma$, the square-integrability of $S$ follows from the existence of the first and second moment of the stopping time $\sigma$. These are given by $E[\sigma]=2\log2$ and $E[\sigma^2]=(2\log 2)^2+8\log 2$, which can be calculated by using the derivatives of the Laplace transform
\be
g(\alpha)=E[\exp(-\alpha \sigma_{a,b})]=\exp(ab-|a|\sqrt{2\alpha+b^2})\quad\text{for $\alpha\geq0$}\label{ex:LT}
\ee
of the stopping times $\sigma_{a,b}:=\inf\{t>0~|~W_t+at=b\}$ at $\alpha=0$ for $a=\frac{1}{2}$ and $b=\log 2$. Then $\lambda=1$ and $K_t=\sigma\wedge t$ and $1$ admits (at least) two FS decompositions $1=1$ with $\widehat\xi^{1}=0\in\Theta$ and $1=\frac{1}{2}+\frac{1}{2}\frac{1}{\E(-W)_\sigma}=\frac{1}{2}+\frac{1}{2}\E(S)_\infty$ with $\widehat\xi^{1}=\frac{1}{2}\E(S)$, where $\frac{1}{2}\E(S)\in\Theta$, as $\frac{1}{2}\E(S)_t=\frac{1}{2}\frac{1}{\E(-W)_{\sigma\wedge t}}\leq 1$ by definition of $\sigma$ and therefore
\begin{align*}
E\left[\frac{1}{4}\left[\E(S)\sint M^S\right]_\infty\right]&=E\left[\frac{1}{4}\int_0^\sigma\E(S)^2_udu\right]\leq E[\sigma],\\
E\left[\left(\int_0^\infty\left|\frac{1}{2}\E(S)_u\right||dA_u^S|\right)^2\right]&=E\left[\left(\int_0^\sigma\frac{1}{2}\E(S)_udu\right)^2\right]\leq E[\sigma^2].
\end{align*}
The linearity of the FS decomposition then implies that no FS decomposition is unique in this market and it therefore only remains to construct a FS decomposition of $K_\infty$. To that end, we define
\begin{align*}
N_t:={}&\frac{E[\E(-W)_\sigma\sigma|\cF_t]}{\E(-W)_{\sigma\wedge t}}=\frac{1}{2}\frac{E[\sigma|\cF_t]}{\E(-W)_{\sigma\wedge t}}\\
={}&\sigma\mathbbm{1}_{\{\sigma\leq t\}}+\frac{1}{2}E[\sigma|\cF_t]\E(S)_t\mathbbm{1}_{\{\sigma\leq t\}}=\sigma\mathbbm{1}_{\{\sigma\leq t\}}+f(S_t,t)\mathbbm{1}_{\{\sigma\leq t\}}
\end{align*}
where $f(s,t):=\frac{1}{2}\big(E[\sigma_{\frac{1}{2},b}]\big|_{b=\log2-(s-\frac{1}{2}t)}+t\big)\exp(s-\frac{1}{2}t)=\big(\log 2-(s-t)\big)\exp(s-\frac{1}{2}t)$ due to the stationary and independent increments of Brownian motion. By Theorem 9 in \cite{S95} and It\^o's formula the so-called generalised FS decomposition of $K_\infty$ is then given by
\begin{align*}
K_\infty=&f(0,0)+\int_0^\sigma \frac{\partial f}{\partial s}(S_u,u)dS_u\\
=&\log 2+\int_0^\sigma\big(\log 2-1-(S_u-u)\big)\exp\left(S_u-\frac{1}{2}u\right)dS_u,
\end{align*}
which coincides with the (classical) FS decomposition, since $\widehat\xi^{K_\infty}:=\frac{\partial f}{\partial s}(S_u,u)\in\Theta$. To see the latter,
we estimate
\begin{align*}
E\left[\left(\int_0^\infty\big|\widehat\xi^{K_\infty}_u\big|d|A^S_u|\right)^2\right]&=E\left[\left(\int_0^\sigma\big|\log 2-1-(S_u-u)\big|du\right)^2\right]\\
&\leq E\left[\left(\int_0^\sigma\big(\log 2-1+|W_u|\big)du\right)^2\right]\\
&\leq 2E\left[\big(\log 2-1\big)^2\sigma^2+\sup_{0\leq u\leq \sigma}|W_u|^2\sigma^2\right]\\
&\leq 2\big((\log 2-1)^2E[\sigma^2]+cE[\sigma^2]E[\sigma^4]\big)
\end{align*}
and
\begin{align*}
E\left[\int_0^\infty\big(\widehat\xi^{K_\infty}_u\big)^2d[M^S]_u\right]&=E\left[\int_0^\sigma\big(\log 2-1-(S_u-u)\big)^2du\right]\\
&\leq 2E\left[\int_0^\sigma\big((\log 2-1)^2+(W_u)^2\big)du\right]\\
&\leq 2E\left[(\log 2-1)^2\sigma+\sup_{0\leq u\leq \sigma}|W_u|^2\sigma\right]\\
&\leq 2\big((\log 2-1)^2E[\sigma]+cE[\sigma^2]E[\sigma^2]\big),
\end{align*}
where we combined the H\"older with the BDG inequality in the last step of each estimate. The moment $E[\sigma^4]$ can again be computed by differentiating the Laplace transform~\eqref{ex:LT}.

Since the strategy $\E(S)\in\TS$ satisfies $\E(S)\sint S_\infty=\E(S)_\infty-1=1$, the solution to the auxiliary problem \eqref{ap-1} is given by $\tvp=\E(S)\in\TS$ and $1\in G_T(\TS):=\{\vt\sint S_T~|~\vt\in\Theta\}$. The latter of course implies that the financial market does not even satisfy the weak no-arbitrage condition of \emph{no approximate profits in $L^2$} that $1\notin \overline{G_T(\TS)}$, where $\overline{\phantom{G}}$ denotes the closure in $L^2(P)$; see Section 4 in \cite{S01}. Plugging $a\E(S)$ in into $U(\cdot)$ with $a>0$, we obtain
$$U(a\E(S))=E[x+(a\E(S))\sint S_\infty]-\frac{\gamma}{2}\Var[x+(a\E(S))\sint S_\infty]=x+a.$$
Therefore the solution to the static MVPS problem \eqref{MVPS} does not exits, as the investor would like to buy more and more stocks exploiting this arbitrage opportunity by sending $a$ to infinity. The solution to the classical Markowitz problem in the formulation \eqref{MP}, however, exists and is given by $\tvt^{(m,x)}=(m-x)\E(S)$. This completes the example.
\end{example}

By Theorem \ref{thmLMVEFS} the LMVE strategy has (in continuous time) the decomposition into the myopically mean-variance efficient (MMVE) strategy and an intertemporal hedging demand consisting of a locally risk minimising strategy as in discrete time, where the MMVE (in continuous time) is defined as follows. 
\begin{defi}
For $\vp,\vt\in\Theta$ and a partition $\tau$ of $[0,T]$, we set
\begin{align}
\widehat{u}^{\tau}[\vp,\vt]&:=\sum_{t_i\in\tau\setminus\{T\}}\frac{U_{t_i}(\vp\mathbbm{1}_{(t_i,t_{i+1}]})-U_{t_i}(\vt\mathbbm{1}_{(t_i,t_{i+1}]})}{E[B_{t_{i+1}}-B_{t_i}|\F_{t_i}]}\mathbbm{1}_{(t_i,t_{i+1}]}\label{dum}\\
&\phantom{:}=\sum_{t_i\in\tau\setminus\{T\}}\frac{\overline{U}_{t_i}(\vp\mathbbm{1}_{(t_i,t_{i+1}]})-\overline{U}_{t_i}(\vt\mathbbm{1}_{(t_i,t_{i+1}]})}{E[B_{t_{i+1}}-B_{t_i}|\F_{t_i}]}\mathbbm{1}_{(t_i,t_{i+1}]}.\nonumber
\end{align}
A strategy $\hvp\in\Theta$ is called \emph{myopically mean-variance efficient (in continuous time)} if
\be
\liminf_{n\to \infty}\widehat u^{\tau_n}[\hvp,\vt]\geq 0 \quad P_B\text{-a.e.}\label{mmvect}
\ee
for any increasing sequence $(\tau_n)_{n\in\N}$ of partitions tending to the identity and any $\vt\in\Theta$.
\end{defi}
With the definition above the MMVE strategy is then given by $\frac{1}{\gamma}\lambda$ as in discrete time.
\begin{prop}\label{prop:MMVE}
There exists a MMVE strategy $\widehat\vp$ if and only if $S$ satisfies (SC) and the terminal value of the MVT process $K_T$ is in $L^2(P)$. In that case, $\widehat\vp$ is unique and given by $\widehat\vp=\frac{1}{\gamma}\lambda$.
\end{prop}
\bp
Since $\widehat{u}^{\tau}[\hvp,\vt]=u^{\tau}[0,\vt]-u^{\tau}[0,\hvp]$, it follows from Lemma \ref{lconvu} that a strategy $\hvp\in\TS$ is MMVE if and only if
\be
\lim_{n\to\infty}\widehat{u}^{\tau_n}[\vp,\vt]=\hvp^\T (c^M \lambda + \eta)-\frac{\gamma}{2}\hvp^\T c^M \hvp-\vt^\T (c^M \lambda + \eta)+\frac{\gamma}{2}\vt^\T c^M \vt\geq0\quad P_B\text{-a.e.}\label{pr:MMVE1}
\ee
for any increasing sequence $(\tau_n)_{n\in\N}$ of partitions tending to the identity and any $\vt\in\Theta$.

Now suppose that $S$ satisfies (SC) and the terminal value of the MVT process $K_T$ is in $L^2(P)$ first. Then the square-integrability of $K_T$ implies that $\frac{1}{\gamma}\lambda\in\TS$, as
$$
K_T=\int_0^T\lambda_u^\T c_u^M\lambda_udB_u=\int_0^T|\lambda_u^\T a_u|dB_u.
$$
Choosing $\hvp=\frac{1}{\gamma}\lambda$ in \eqref{pr:MMVE1} and completing squares gives that
$\frac{\gamma}{2}\left(\vt-\hvp\right)^\T c^M\left(\vt-\hvp\right)\geq0$ $P_B$-a.e.~for all $\vt\in\TS$ and therefore that $\hvp=\frac{1}{\gamma}\lambda$ is MMVE.

Conversely, assume that there exists a MMVE strategy $\hvp$. Then plugging in the strategies $\vt=\hvp+\eta\mathbbm{1}_{D_k}\in\TS$ with $D_k=\{|\eta^\T a|\leq k\}$ and $\vt=\frac{1}{\gamma}\lambda\mathbbm{1}_{D_k}\in\TS$ with $D_k=\{\lambda^\T c^M\lambda^\T +|\lambda^\T a|\leq k\}$ for $k\in\N$ into \eqref{pr:MMVE1} gives that $-\eta^\T\eta\geq 0$ $P_B$-a.e.~and $\frac{\gamma}{2}(\frac{1}{\gamma}\lambda-\hvp)^\T c^M(\frac{1}{\gamma}\lambda-\hvp)\geq0$ $P_B$-a.e.~on $D_k$. Therefore choosing   $k$ sufficiently large implies that $S$ satisfies (SC), i.e.~$\eta=0$, and that $\hvp=\frac{1}{\gamma}\lambda$, as we would otherwise derive a contradiction. Since $\hvp\in\TS$, we also obtain the square-integrability of $K_T$ from the latter by $K_T=\gamma \int_0^T\hvp_u^\T a_udB_u\in L^2(P)$, which completes the proof.
\ep
Instead of optimising the conditional mean-variance criterion as the MMVE investor in each step separately the LMVE investor seeks to invest more sustainably by taking also the investment horizon $T$ into account. So the difference between the LMVE and MMVE strategy is that the LMVE investor hedges in addition to holding the MMVE strategy the risk coming from considering this strategy not only over the next period but on the entire remaining time interval. The risk induced by this is driven by the stochastic investment opportunity set and can be represented by $\frac{1}{\gamma}K_T$ using the MVT. This risk is then minimised by the LMVE investor in the sense of local risk minimisation which yields the additional intertemporal hedging demand $\frac{1}{\gamma}\widehat\xi=\widehat\xi(\hvt)$ in the LMVE strategy. In fact the LMVE and the MMVE strategy coincide and the relations in Theorem \ref{thmLMVEFS} simplify, if the investment opportunity set or more generally the terminal value of the MVT process $K_T$ is deterministic. Note, however, that the optimal strategy $\tvt$ for the static MVPS problem \eqref{MVPS} is still different. The price processes $S$ has a deterministic investment opportunity set, if it has independent increments, which is for example the case if $S$ is a L\'evy process or the exponential of one. A discussion and comparison between the LMVE/MMVE strategy and the solution to the static MVPS problem \eqref{MVPS} in the Black-Scholes model is given in Section 1.4 in \cite{BC}.
\begin{cor}\label{cordos}
Suppose that $S$ satisfies (SC) and that the terminal value of the MVT process $K_T$ is deterministic. Then:
Suppose that $S$ satisfies (SC) and that the terminal value of the MVT process $K_T$ is deterministic. Then:
\bi
\item[1)] The FS-decomposition of $K_T$ reduces to $K_T=\widehat K_0$, the LMVE strategy $\hvt$ exists and coincides with the MMVE, i.e.~$\hvt=\hvp=\frac{1}{\gamma}\lambda$, and the equations in Theorem \ref{thmLMVEFS} simplify to $Z_t(\hvt)=\frac{1}{\gamma}(K_T-K_t)$ and
$$U_t(\widehat \vt)=U_t(\widehat \vp)=x+\int_0^t\frac{1}{\gamma}\lambda dS+\frac{1}{2\gamma}(K_T+K_t).$$  
\item[2)] If $S$ is in addition continuous or the entire MVT process $K$ is deterministic, the optimal strategy $\tvt$ for the static MVPS problem \eqref{MVPS} is given by
$$\tvt=\frac{1}{\gamma}\frac{1}{\E(-\widetilde K)_T}\E(-\widetilde\lambda\sint S)_-\widetilde\lambda,$$
where $\widetilde\lambda:=\frac{\lambda}{1+\Delta K}$ and $\widetilde K_t:=\int_0^t \frac{1}{1+\Delta K_u}dK_u$, and
$$U_t(\tvt)=\frac{1}{\gamma}\E(-\widetilde\lambda\sint S)_t\frac{\E(-\widetilde K)_T}{\E(-\widetilde K)_t}\left(1-\frac{1}{2}\E(-\widetilde\lambda\sint S)_t\left(1-\frac{\E(-\widetilde K)_T}{\E(-\widetilde K)_t}\right)\right).$$
\ei
\end{cor}
\bp
1) Since $K_T\in L^\infty(P)$, we have that $\E(-\lambda\sint M)$ is regular and satisfies $R_2(P)$. Therefore the FS decomposition of $K_T$ exists, is unique and given by $K_T=\widehat{K}_0$. The assertions on the LMVE and MMVE strategy follow then from Corollary \ref{ces}, Theorem~\ref{thmLMVEFS} and Proposition~\ref{prop:MMVE}.

2) Under these assumptions the solution $\tvp$ to \eqref{ap-1} is given by $\tvp=\E(-\lambda\sint S)_-\lambda$ by Theorem 7 and 8 in \cite{S95}. The formulas for $\tvt$ and $U(\tvt)$ then follow by \eqref{gds} and direct computations. This completes the proof.
\ep

The optimality condition \eqref{optcond} basically tells us that the locally mean-variance efficient strategy $\widehat\vt$ is a fixed point of the mapping
$\widehat J: \Theta\to \Theta$ given by
\be
\widehat J(\vt)=\frac{1}{\gamma}\lambda-\xi(\vt).\label{defpsi}
\ee
Exploiting again the relation to the FS decomposition, we can show that this fixed point can be obtained by an iteration.  Since the iteration algorithm reduces to a backward recursion in discrete time, this can be seen as a continuous-time analogue of the recursive derivation of the LMVE strategy in Lemma \ref{lrrd} in discrete time. Moreover, the characterisation of the LMVE strategy as a fixed point illustrates the game-theoretic interpretation of the optimal strategy as an equilibrium of an intrapersonal game.

\bl
If the mean-variance tradeoff process $K$ is bounded and continuous, the mapping $\widehat J(\vt)=\frac{1}{\gamma}\lambda-\xi(\vt)$ is a contraction on $(\Theta,\|.\|_{\beta,\infty})$ with modulus of contraction $c\in(0,1)$ where
$$\|\vt\|_{\beta,\infty}:=\left\|\left(\int_0^T\frac{1}{\E(-\beta K)_u} \vt_u^\T d\la M\ra_u\vt_u\right)^{\frac{1}{2}}\right\|_{\LiiP}.$$
In particular, the locally mean-variance efficient strategy $\widehat\vt$ is  unique and  given as the limit
$$\widehat\vt=\lim_{n\to\infty}\vt^n$$
in $(\Theta,\|.\|_{\beta,\infty})$, where $\vt^{n+1}=\widehat J(\vt^n)$ for $n\geq 1$, for any starting value $\vt^0=\vt\in\Theta$.
\el
\bp
Integrating both sides of \eqref{defpsi} with respect to $M$ and using the definition of $\xi(\vt)$ we obtain
$$\int_0^T\widehat J_u(\vt)dM_u=\int_0^T\frac{1}{\gamma}\lambda_u dM_u+Y_0(\vt)+L_T(\vt)-\int_0^T\vt_u dA_u$$
and from this
$$\frac{1}{\gamma}K_T-\int_0^T\left(\frac{1}{\gamma}\lambda_u-\vt_u\right) dA_u=Y_0(\vt)+\int_0^T\left(\frac{1}{\gamma}\lambda_u-\widehat J_u(\vt)\right) dM_u+L_T(\vt)$$
after rearranging terms and inserting the zero term $\frac{1}{\gamma}K_T-\frac{1}{\gamma}\int_0^T\lambda_u dA_u$. Comparing the last equation with the definition of the mapping $J$ in the proof of Corollary 5 in \cite{PRS98} gives that $\widehat J(\vt)=\frac{1}{\gamma}\lambda-J\left(\frac{1}{\gamma}\lambda-\vt\right)$, as $L(\vt)$ is strongly orthogonal to $M$ and therefore the right-hand side is the GKW decomposition of the left-hand side. If $K$ is bounded and continuous, it follows from the arguments in the proof of Corollary 5 in \cite{PRS98} that \mbox{$J:(\Theta,\|.\|_{\beta,\infty})\to (\Theta,\|.\|_{\beta,\infty})$}, and hence also $\widehat J$, is a contraction with modulus of contraction $c\in(0,1)$, which immediately implies that the sequence $(\vt^n)$ converges to $\widehat\vt$ for any starting value $\vt^0=\vt\in\Theta$ by Banach's fixed point theorem.
\ep
\br
\bi
\item[\textbf{1)}] Note that this proves that in our setting, the LMVE strategy $\widehat\vt$ can indeed be obtained by the iteration procedure suggested in \cite{BM}.

\item[\textbf{2)}] If the jumps of $K$ are uniformly bounded by some constant $b\in(0,1)$, it follows from the remark following Corollary 5 in \cite{PRS98} that $J$ and therefore $\widehat J$ are still contractions on $(\Theta,\|.\|_{\beta,\infty})$ with modulus of contraction $c\in(0,1)$; see also Lemma \ref{lJ} later.

\item[\textbf{3)}] Using the ``salami technique'' in \cite{MS95}, one can show that the iterations still converge if $K$ is only bounded, even though the modulus of contraction $c$ is then not \mbox{necessarily in $(0,1)$.}
\ei
\er
\section{Convergence of solutions}
To establish a link between the intuitive situation in discrete time, where the time-consistent optimal strategy is found by a backward recursion, and the continuous-time formulation given by a limit, we show that the solutions obtained in discretisations of a continuous-time model converge to the solution in continuous time. This underlines that our formulation in continuous time is indeed the natural extension of that in discrete time. For this result, however, we need to discretise in an appropriate sense.

Let $(\tau_n)_{n\in\N}$ be an increasing sequence of partitions of $[0,T]$ such that $|\tau_n|\to 0$ and assume for simplicity that $S$ is one dimensional, i.e.~$d=1$. Then we choose $B=\la M \ra$ and set $P_B=P_{\la M \ra}$ which we deliberately denote by $P_M$ in this section. Moreover, we denote by $S^n$ the RCLL discretisation of $S$ with respect to the partition $\tau_n$, which is given by $S^n_{t_i}=S_{t_i}$ for all $t_i\in\tau_n$ and constant on $[t_i,t_{i+1})$, and by $\mathbb{F}^n=(\cF^n_t)_{0\leq t\leq T}$ the filtration given by $\cF^n_t=\cF_{t_i}$ for $t\in[t_i,t_{i+1})$. This discretisation corresponds to the situation that we only trade at a finite number of given trading dates $t_i\in\tau_n$. Under the assumption that $S=S_0+M+A$ is square-integrable, all $S^n$ are square-integrable semimartingales on $(\Omega,\cF,\mathbb{F}^n,P)$ with Doob decompositions $S^n=S_0+\bar M^n+ \bar A^n$ in $\FF^n$ as constructed in Section \ref{sec:dt}. Since the processes $\bar M^n$ and $\bar A^n$ are a priori only defined on $\tau_n$, we extend them to piecewise constant right-continuous processes on $[0,T]$ by taking $\bar M^n_t=\bar M^n_{t_i}$ and $\bar A^n_t=\bar A^n_{t_i}$ for $t\in[t_i,t_{i+1})$ and $t_i\in\tau_n$, which is consistent with the Doob--Meyer decomposition of the semimartingale $S^n$ with respect to the filtration $\FF^n$. This will be the usual embedding we use to include the discrete-time case into the continuous-time framework (as for example explained in Sections I.1f and I.4g in \cite{JS}). Note that $\bar M^n$ and $\bar A^n$ are not obtained by discretising the continuous-time processes $M$ and $A$ in the same way as we obtain $S^n$ from $S$; this explains the choice of notation, and it is the source of the difficulties in proving our result. For later references we denote by $\cM_0^2(P,\FF^n)$ the space of all square-integrable $\FF^n$-martingales null at zero and by $\cH^2(P,\FF^n)=\cH^2(\FF^n)$ the space of all special $\FF^n$-semimartingales with finite $\cH^2(\FF^n)$-norm.

To ensure the existence of a solution in the continuous-time setting, we assume the conditions of Corollary \ref{ces}. These also yield the existence of solutions in all discretised settings, in which we have
$$
\lambda^n=\sum_{t_{i+1}\in\tau_n\setminus\{T\}}\frac{\Delta \bar A^n_{t_{i+1}}}{E[(\Delta\bar M^n_{t_{i+1}})^2 |\cF_{t_i}]}\mathbbm{1}_{(t_i,t_{i+1}]}$$
and
$$K^n_T=\sum_{t_{i+1}\in\tau_n\setminus\{T\}}\frac{\Delta \bar A^n_{t_{i+1}}}{E[(\Delta\bar M^n_{t_{i+1}})^2 |\cF_{t_i}]}\Delta \bar A^n_{t_{i+1}}.$$
Since we are changing our optimisation criterion each time we increase the partition, we cannot use the elegant approximation techniques for standard utility maximisation problems as in \cite{KP08} to obtain the convergence of the solutions. Instead, we have to work directly with the structure of the solution. We exploit that we have $\hvt^n=\frac{1}{\gamma}(\lambda^n-\widehat\xi^n)$ and $\hvt=\frac{1}{\gamma}(\lambda-\widehat\xi)$ as global descriptions in discrete as well as in continuous time, where $\widehat\xi^n$ is the integrand in the discrete-time F\"ollmer--Schweizer decomposition of $K^n_T$ with respect to $S^n$ and $(\Omega,\cF,\mathbb{F}^n,P)$, i.e.
$$K_T^n=\widehat K^n_0+\int_0^T\widehat \xi^n_u dS^n_{u}+\widehat L^n_T=\widehat  K^n_0+\sum_{t_i\in\tau_n\setminus\{0\}}\widehat \xi_{t_i}^n\Delta S^n_{t_i}+\widehat  L^n_T$$
for $n\in\N$, and $\widehat \xi$ is the integrand in the continuous-time F\"ollmer--Schweizer decomposition of $K_T$ with respect to $S$, i.e.
$$K_T=\widehat K_0+\int_0^T\widehat \xi_u dS_{u}+\widehat L_T.$$
For the proof of the convergence $\hvt^n=\frac{1}{\gamma}(\lambda^n-\widehat\xi^n)\overset{L^2(M)}{\longrightarrow}\hvt=\frac{1}{\gamma}(\lambda-\widehat\xi)$ we then show that
\be
\lambda^n\overset{L^2(M)}{\longrightarrow}\lambda^\infty:=\lambda\label{cln}
\ee
and
\be
\widehat\xi^n\overset{L^2(M)}{\longrightarrow}\widehat\xi^\infty:=\widehat\xi\label{cxin}
\ee
separately. For the latter we also need to establish that
\be
K^n_T\overset{L^2(P)}{\longrightarrow}K^{\infty}_T:=K_T.\label{cKn}
\ee

The main difficulty is that the canonical decomposition is not stable under discretisation in the following sense. As already pointed out, $\bar M^n$ and $\bar A^n$ are not simply obtained by discretising $M$ and $A$ to $M^n_t:=M_{t_i}$ and $A^n_t:=A_{t_i}$ for $t\in[t_i,t_{i+1})$. From the discrete-time Doob decomposition, they are rather given by the processes $\bar M_t^n:=M_t^n+ M_t^{A,n}$, where $M^{A,n}_t:=\sum_{k=1}^i(\Delta A^n_{t_k}-E[\Delta A^n_{t_k}|\cF_{t_{k-1}}])$, and $\bar A_t^n:=\sum_{k=1}^iE[\Delta A^n_{t_k}|\cF_{t_{k-1}}]$ for $t\in[t_i,t_{i+1})$. Note that we deliberately set $\la M^n \ra:=\la M^n \ra^{\FF^n}$, $\la \bar M^n \ra:=\la \bar M^n \ra^{\FF^n}$ and $\la M^{A,n} \ra:=\la M^{A,n} \ra^{\FF^n}$ to simplify notation. For the $\mathbb{F}^n$-martingale $M^{A,n}$, which represents the ``discretisation error'' in the canonical decomposition, we already know from Lemma \ref{lA_3} that
$$\lim_{n\to\infty}\frac{d\la M^{A,n}\ra}{d\la M^{n}\ra}=\lim_{n\to\infty}\sum_{t_{i}\in\tau_n\setminus\{0\}}\frac{\Var\left[A_{t_i}-A_{t_{i-1}}|\F_{t_{i-1}}\right]}{E[\la M\ra _{t_{i}}-\la M \ra_{t_{i-1}}|\F_{t_{i-1}}]}\mathbbm{1}_{(t_{i-1},t_{i}]}=0\qquad P_M\text{-a.e.}$$
Moreover, if $\lambda\sint M\in bmo_2$, we have
\begin{align}
\Var\left[A_{t_i}-A_{t_{i-1}}|\F_{t_{i-1}}\right]&\leq E\big[(A_{t_i}-A_{t_{i-1}})^2\big|\F_{t_{i-1}}\big]\nonumber\\
&=E\left[\left(\int_{t_i}^{t_{i-1}}\lambda_ud\la M\ra_u\right)^2\bigg|\F_{t_{i-1}}\right]\nonumber\\
&\leq E\left[\left(\int_{t_i}^{t_{i-1}}\lambda_u^2d\la M\ra_u\right)\left(\int_{t_i}^{t_{i-1}}d\la M\ra_u\right)\bigg|\F_{t_{i-1}}\right]\nonumber\\
&\leq \big\|(\mathbbm{1}_{(t_{i-1},t_{i}]}\lambda)\sint M\big\|_{bmo_2}^2E\left[\int_{t_{i-1}}^{t_i}d\la M\ra_u\bigg|\F_{t_{i-1}}\right]\label{Abmo}
\end{align}
by applying Jensen's inequality and the definition of the $bmo_2$-norm, which gives
\be
\left\|\frac{d\la M^{A,n}\ra}{d\la M^{n}\ra}\right\|_{L^\infty(P_M)}\leq \sup_{t_{i}\in\tau_n\setminus\{0\}}\big\|(\mathbbm{1}_{(t_{i-1},t_{i}]}\lambda)\sint M\big\|_{bmo_2}^2\leq\big\|\lambda\sint M\big\|_{bmo_2}^2\label{eMAn}.
\ee
However, to obtain the convergences \eqref{cln}--\eqref{cKn} above, we shall finally need to use that $\frac{d\la M^{A,n}\ra}{d\la M^{n}\ra}\longrightarrow0$ in $L^\infty(P_M)$, and we also need a tight control in $L^\infty(P_M)$ on the $K^n_T$ and on $(\Delta K^n)^*_T:=\sup_{0\leq s\leq T}|\Delta K^n_s|$ in $L^\infty(P)$, for an arbitrary increasing sequence of partitions tending to the identity. A sufficient condition for this is given in the following lemma.
\bl\label{lKn}
Assume that $K=\int\mu^K dt$ and that $\mu^K$ is uniformly bounded in $\om$ and $t$ by some constant $c_\mu>0$. Then:
\bi
\item[1)] $\frac{d\la M^{A,n}\ra}{d\la M^{n}\ra}\overset{L^\infty(P_M)}{\longrightarrow}0$, which implies $\frac{d\la\bar M^{n}\ra}{d\la M^{n}\ra}\overset{L^\infty(P_M)}{\longrightarrow}1$ and $\frac{d\la M^{n}\ra}{d\la\bar M^{n}\ra}\overset{L^\infty(P_M)}{\longrightarrow}1$.
\item[2)] There exist $n_0\in\N$ and $b\in(0,1)$ such that $\sup_{n\geq n_0}\|K^n_T\|_{L^\infty(P)}$ is finite and $\sup_{n\geq n_0}\|(\Delta K^n)^*_T\|_{L^\infty(P)}\leq b$, and moreover $(\Delta K^n)^*_T\to0$ in $L^\infty(P)$.
\ei
\el
\bp
1) This immediately follows from \eqref{eMAn} above and observing that
$$\| (\lambda\mathbbm{1}_{(s,t]})\sint M\|^2_{bmo_2}\leq\sup_{s\leq u\leq t}\|E[K_t-K_u|\cF_u]\|_{L^\infty(P)}\leq c_\mu(t-s).$$
From $\frac{d\la M^{A,n}\ra}{d\la M^{n}\ra}\overset{L^\infty(P_M)}{\longrightarrow}0$ we then obtain that $\frac{d\la\bar M^{n}\ra}{d\la M^{n}\ra}\overset{L^\infty(P_M)}{\longrightarrow}1$ by using $\bar M^n=M^n+M^{A,n}$ and the Cauchy--Schwarz inequality. The latter convergence also implies that $\frac{d\la M^{n}\ra}{d\la\bar M^{n}\ra}\overset{L^\infty(P_M)}{\longrightarrow}1$.\\
2) Since $\frac{d\la M^{n}\ra}{d\la\bar M^{n}\ra}\overset{L^\infty(P_M)}{\longrightarrow}1$, we can choose $n_0\in\N$ such that $\sup_{n\geq n_0}\big\|\frac{d\la M^{n}\ra}{d\la\bar M^{n}\ra}\big\|_{L^\infty(P_M)}\leq c$ for some $c>0$. By the Cauchy--Schwarz inequality we can estimate
$$(\Delta \bar A^n_{t_{i+1}})^2=\left(E\left[\mbox{$\int_{t_{i}}^{t_{i+1}}\lambda_u d\la M \ra_u$}\Big|\cF_{t_i}\right]\right)^2\leq E[K_{t_{i+1}}-K_{t_i}|\cF_{t_i}]E\left[\mbox{$\int_{t_{i}}^{t_{i+1}}d\la M \ra_u$}\Big|\cF_{t_i}\right],$$
which gives for $n\geq n_0$ that
\begin{align*}
\|(\Delta K^n)^*_T\|_{L^\infty(P)}&=\left\|\sup_{t_{i+1}\in\tau_n\setminus\{T\}}\frac{(\Delta \bar A^n_{t_{i+1}})^2}{E[(\Delta\bar M^n_{t_{i+1}})^2 |\cF_{t_i}]}\right\|_{L^\infty(P)}\\
&\leq\left\|\frac{d\la M^{n}\ra}{d\la\bar M^{n}\ra}\right\|_{L^\infty(P_M)}\left\|\sup_{t_{i+1}\in\tau_n\setminus\{T\}}E[K_{t_{i+1}}-K_{t_i}|\cF_{t_i}]\right\|_{L^\infty(P)}\leq c_\mu c|\tau_n|\overset{n\to\infty}{\longrightarrow} 0.
\end{align*}
By the same arguments we obtain $\|\Delta K^n_{t_{i+1}}\|_{L^\infty(P)}\leq c_\mu c(t_{i+1}-t_i)$ for $n\geq n_0$ and therefore $\sup_{n\geq n_0}\|K^n_T\|_{L^\infty(P)}\leq c_\mu cT$ after summing up. This completes the proof.
\ep
Because $\frac{d\la M^{n}\ra}{d\la\bar M^{n}\ra}\overset{L^\infty(P_M)}{\longrightarrow}1$ implies the existence of some $n_0\in\N$ and $c>0$ such that $\sup_{n\geq n_0}\big\|\frac{d\la M^{n}\ra}{d\la\bar M^{n}\ra}\big\|_{L^\infty(P_M)}\leq c$, we can already prove \eqref{cln} via the next lemma.
\bl\label{lconvMtoA}
Let $\lambda\in L^2(M)$ and assume that $\left\|\frac{d\la M^{n}\ra}{d\la\bar M^{n}\ra}\right\|_{L^\infty(P_M)}\leq c$ for some $c>0$. Then $\lambda^n\overset{L^2(M)}{\longrightarrow}\lambda.$
\el
\bp
Using (SC), we can write
\begin{align*}
\lambda^n=\sum_{t_{i+1}\in\tau_n\setminus\{T\}}\frac{E[\mbox{$\int_{t_{i}}^{t_{i+1}}\lambda_u d\la M \ra_u$}|\cF_{t_i}]}{E[\mbox{$\int_{t_{i}}^{t_{i+1}}d\la M \ra_u$}|\cF_{t_i}]}\frac{E[(\Delta M^n_{t_{i+1}})^2 |\cF_{t_i}]}{E[(\Delta\bar M^n_{t_{i+1}})^2 |\cF_{t_i}]}\mathbbm{1}_{(t_i,t_{i+1}]}=E_M\big[\, \lambda\,  \big|\, \cP^{\tau_n}\, \big]\frac{d\la M^{n}\ra}{d\la \bar M^{n}\ra}.
\end{align*}
Since the $\sigma$-fields $\cP^{\tau_n}$ increase to the predictable $\sigma$-field $\cP$ and $\lambda\in L^2(P_M)$ is predictable, $\big(E_M[\, \lambda\, |\, \cP^{\tau_n}\, ]\big)_{n\in\N}$ is a square-integrable martingale on $\big(\OmT, \cP, (\cP^{\tau_n})_{n\in\N},P_M\big)$ which converges to $\lambda$ $P_M$-a.e.~and in $L^2(P_M)$ by the martingale convergence theorem. To conclude the assertion, we use the following simple fact with $X^n=\lambda ^n$, $Y^n=\frac{d\la M^{n}\ra}{d\la\bar M^{n}\ra}$ and $P=P_M$. Let $(X^n)$ and $(Y^n)$ be two sequences of random variables such that $X^n\to X$ $P$-a.s.~and in $L^2(P)$, $Y^n\to Y$ $P$-a.s.~and $\|Y^n\|_{L^\infty(P)}\leq c$ and $\|Y\|_{L^\infty(P)}\leq c$ for some $c>0$. Then $X^nY^n\to XY$ $P$-a.s.~and in $L^2(P)$. Due to the estimate
\begin{align*}
\|X^nY^n-XY\|_{L^2(P)}&\leq\|(X^n-X)Y^n\|_{L^2(P)}+\|X(Y^n-Y)\|_{L^2(P)}\\
&\leq c \|X^n-X\|_{L^2(P)}  +2c\|X\|_{L^2(P)}
\end{align*}
this can be seen by using that $X^nY^n\to XY$ $P$-a.s.~and Lebesgue's dominated convergence with majorant $2c|X|\in L^2(P)$, which completes the proof.
\ep
For the proof of \eqref{cKn} we establish the following result which is slightly more general than we actually need.
\bl
Let $\lambda\sint M\in bmo_2$ and assume that $\xi^n\overset{L^2(M)}{\longrightarrow}\xi$ and that $\xi^n$ is $\cP^{\tau_n}$-measurable for each $n\in\N$. Then $\xi^n\sint\bar A^{n}_T\to\xi\sint A_T$ in $L^2(P)$. 
\el
\bp
As each $\xi^n$ is piecewise constant along $\tau_n$, we obtain
\begin{align*}
E\left[\left(\xi^n\sint\bar A^{n}_T-\xi\sint A_T\right)^2\right]&=E\left[\left(\sum_{t_i\in\tau_n\setminus\{0\}}\xi^n_{t_i}(\Delta\bar A^{n}_{t_i}-\Delta A^n_{t_i})+(\xi^n-\xi)\sint A_T\right)^2\right]\\
&=E\left[\left(\sum_{t_i\in\tau_n\setminus\{0\}}-\xi^n_{t_{i}}\Delta M^{A,n}_{t_i}-(\xi^n-\xi)\sint A_T\right)^2\right]\\
&\leq2E\left[\left(\xi^n\sint M^{A,n}_T\right)^2\right]+2E\left[\big((\xi^n-\xi)\sint A_T\big)^2\right]
\end{align*}
and therefore that
\begin{align}
E\left[\left(\xi^n\sint \bar A^{n}_T-\xi\sint A_T\right)^2\right]&\leq 2E\big[(\xi^n)^2\sint\la M^{A,n}\ra_T\big]+2\|\xi^n-\xi\|^2_{L^2(A)}\label{eq1}
\end{align}
by It\^o's isometry, since $\xi^n\sint M^{A,n}\in\cM^2_{0}(P,\FF^{n})$. Replacing $\la M^{A,n}\ra$ by $\frac{d\la M^{A,n}\ra}{d\la M^{n}\ra}\sint \la M^{n}\ra$ and using that $\xi^n\in L^2(M)$ and $\frac{d\la M^{A,n}\ra}{d\la M^{n}\ra}$ are piecewise constant along $\tau_n$, we can write
$$E\big[(\xi^n)^2\sint\la M^{A,n}\ra_T\big]=E\left[\left(\xi^n\sqrt{\frac{d\la M^{A,n}\ra}{d\la M^{n}\ra}}\right)^2\sint\la M\ra_T\right]=E_M\left[\left(\xi^n\sqrt{\frac{d\la M^{A,n}\ra}{d\la M^{n}\ra}}\right)^2\right].$$
Moreover, $\left(\frac{d\la M^{A,n}\ra}{d\la M^{n}\ra}\right)_{n\in\N}$ is bounded in $L^\infty(P_M)$ due to \eqref{eMAn}. Applying again the simple fact from the proof of the previous lemma, this time with $X^n=\xi^n$, $Y^n=\sqrt{\frac{d\la M^{A,n}\ra}{d\la M^{n}\ra}}$ and $P=P_M$, we obtain that $\left(\xi^n\sqrt{\frac{d\la M^{A,n}\ra}{d\la M^{n}\ra}}\right)$ converges to $0$ in $L^2(P_M)$. To complete the proof we observe that the second term on the right-hand side of \eqref{eq1} also vanishes, since we have $\|\xi^n-\xi\|^2_{L^2(A)}\leq8\|\lambda\cdot M\|_{bmo_2}\big\|\xi^n-\xi\big\|^2_{L^2(M)}$ by Theorem 3.3 in \cite{DMSSS97}. By combining Jensen's inequality with the definition of the $bmo_2$-norm as in the last line of \eqref{Abmo}, we can replace the constant $8$ actually by $1$.
\ep
Now \eqref{cKn} follows immediately by combining the two previous lemmas.
\begin{cor}\label{corKn}
Let $\lambda\cdot M\in bmo_2$ and assume that $\left\|\frac{d\la M^{n}\ra}{d\la\bar M^{n}\ra}\right\|_{L^\infty(P_M)}\leq c$ for some $c>0$. Then $K^n_T\overset{L^2(P)}{\longrightarrow} K_T$.
\end{cor}
To conclude the convergence of the LMVE strategies, it then remains to show \eqref{cxin}.  For this we establish the convergence of the discrete F\"ollmer--Schweizer decompositions obtained in a sequence of discretisations of a financial market as the partitions tend to the identity. More precisely, we want to prove the following result.
\bt\label{mthmconv}
Suppose that $K$ is bounded, $\frac{d\la M^{A,n}\ra}{d\la M^{n}\ra}\overset{L^\infty(P_M)}{\longrightarrow}0$ and that there exist $n_0\in\N$ and $b\in(0,1)$ such that $\sup_{n\geq n_0}\|K^n_T\|_{L^\infty(P)}<\infty$ and $\sup_{n\geq n_0}\|(\Delta K^n)^*_T\|_{L^\infty(P)}\leq b$. Let $H^n, H\in\LiiP$ be contingent claims and $(\tau_n)_{n\in\N}$ an increasing sequence of partitions of $[0,T]$. Write the F\"ollmer--Schweizer decompositions of $H^n$ and $H$ with respect to $S^n$ on $(\Om,\F,\FF^n,P)$ and $S$ on $(\Om,\F,\FF,P)$ as
\be
H^n=\widehat H^n_0+\int_0^T\widehat \xi^n_u dS^n_{u}+\widehat L^n_T=\widehat H^n_0+\sum_{t_i\in\tau_n\setminus\{0\}}\widehat\xi_{t_i}^n\Delta S^n_{t_i}+\widehat L^n_T\label{FSn}
\ee
and
\be
H=\widehat H_0+\int_0^T\widehat \xi_u dS_{u}+\widehat L_T\label{FS}.
\ee
Then $\widehat \xi^n$ converges to $\widehat\xi$ in $L^2(P_M)$, if $H^n\to H$ in $\LiiP$ and $|\tau_n|\to 0$.
\et
For the rest of the section, we always work under the assumptions of Theorem \ref{mthmconv}. To simplify notation we set $H^\infty:=H$, $S^\infty:=S$, $\widehat\xi^\infty:=\widehat\xi$, $\bar M^\infty:=M^\infty=M$, $\bar A^\infty:=A$, $K^\infty:=K$ etc. Note that $M^{A,\infty}=0$. As we deal with GKW decompositions with respect to different martingales, we denote the GKW decomposition of a random variable $H\in L^2(P)$ with respect to $X\in\cM^2_{0}(P,\FF^n)$ for some $n\in\overline{\N}:=\N\cup\{+\infty\}$ by
$$H=E[H|\cF_0]+\int_0^T\xi_u(X,H)dX_u+L_T(X,H),$$
if we need to clarify the dependence on $H$ and $X$. If $n\in\N$, i.e.~in discrete time, we have
$$\xi_t(X,H)=\frac{E\big[H\Delta X_{t_i}\big|\cF_{t_{i-1}}\big]}{E\big[(\Delta X_{t_i})^2\big|\cF_{t_{i-1}}\big]}$$
for $t\in[t_i,t_{i+1})$. The first step in the proof of Theorem \ref{mthmconv} is then to observe that the F\"ollmer--Schweizer decomposition can be obtained under our assumptions by a fixed point iteration, as is shown in Lemma \ref{lJ} below. This is basically the proof of Corollary 5 in \cite{PRS98} and the remark following that. However, as we are interested in the convergence of different F\"ollmer--Schweizer decompositions, we need to establish that several constants are independent of $n$. This allows us to adapt the method of proof of \cite{BDM01} and \cite{BDM02} to our situation. That method is used there to show the convergence of solutions to discretisations of a continuous-time BSDE to the solution in continuous time. Denoting by $\xi^{n,p}$ the $p$-th step of the fixed point iteration leading to $\widehat \xi^n$, for $n\in\overline{\N}$, (where $\widehat\xi^{\infty}=\widehat\xi$) gives the decomposition
$$\widehat \xi^n-\widehat\xi=(\widehat \xi^n-\xi^{n,p})+(\xi^{n,p}-\xi^{\infty,p})+(\xi^{\infty,p}-\widehat\xi).$$
To establish the convergence of the FS decompositions, it then remains to show that $\xi^{n,p}$ converges to $\widehat \xi^n$ in $L^2(M)$ for sufficiently large $n$ uniformly in $n$ as $p\to\infty$, and that $\xi^{n,p}$ converges to $\xi^{\infty,p}$ in $L^2(M)$ for each $p\in\N_0$ as $n\to\infty$, which will be done in Propositions \ref{p1} and \ref{p2}.
\bl\label{lJ}
Under the assumptions of Theorem \ref{mthmconv} there exist $n_0\in\N$ and $b\in(0,1)$ such that the following hold for all $n\in\overline{\N}_{\geq n_0}:=\{n\in \overline{\N}~|~n\geq n_0\}$:
\bi
\item[1)] $\Theta_{S^n}=L^2(\bar M^n)$, and
$$\|\vt\|_{\beta,n}:=\left\|\left(\int_0^T\frac{1}{\E(-\beta K^n)_u} \vt_u d\la \bar{M}^n\ra_u\vt_u\right)^{\frac{1}{2}}\right\|_{\LiiP}$$
defines a norm on $\Theta_{S^n}$ which is equivalent to $\|.\|_{L^2(\bar M^n)}$ for any $\beta\in(0,\frac{1}{b})$, where the equivalence constant $k$ can be chosen independent of $n$, e.g.
$$k=\max\left( \exp\left(\frac{\beta}{1-\beta b}\sup_{n\geq n_0}\|K^n_T\|_{L^\infty(P)}\right), \left\|\frac{1}{\E(-\beta K^\infty)}_T\right\|_{L^\infty(P)}\right).$$
\item[2)] The mapping $J^n:\Theta_{S^n}\to \Theta_{S^n}$ which maps $\vt\in\Theta_{S^n}$ into the integrand
$$\xi\left(\bar M^n,H^n-\int_0^T\vt_ud\bar A^n_u\right)$$
of $\bar M^n$ in the GKW decomposition of $H^n(\vt):=H^n-\int_0^T\vt_ud\bar A^n_u$, i.e.
$$H^n(\vt)=E\left[H^n(\vt)|\cF_0\right]+\int_0^T\xi_u(\bar M^n,H^n(\vt)) d\bar{M}^n_u+L^n_T(\bar M^n,H^n(\vt)),$$
is a contraction on $(\Theta_{S^n},\|.\|_{\beta, n})$ with a modulus of contraction $c\in(0,1)$ that can be chosen independent of $n$, for any $\beta\in(1,\frac{1}{b})$.
\item[3)] The integrand $\widehat\xi^{n}$ in the F\"ollmer--Schweizer decomposition is given as the limit
$$\widehat\xi^{n}=\xi^{n,\infty}=\lim_{p\to\infty}\xi^{n,p}$$
in $(\Theta_{S^n},\|.\|_{\beta, n})$, where $\xi^{n,0}=0$ and $\xi^{n,p}=J^n(\xi^{n,p-1})$ for all $p\in\N$.
\ei
\el
\bp
1) Under the assumptions of Theorem \ref{mthmconv}, there exists $n_0\in\N$ with
$$\sup_{n\in\overline{\N}_{\geq n_0}}\|K^n_T\|_{L^\infty(P)}<\infty$$
and therefore
$$\|\vt\|_{L^2(\bar M^n)}\leq \|\vt\|_{\Theta_{S^n}}\leq \big(1+\sup_{n\in\overline{\N}_{\geq n_0}}\|K^n_T\|_{L^\infty(P)}^{\frac{1}{2}}\big)\|\vt\|_{L^2(\bar M^n)},$$
which implies that $\Theta_{S^n}=L^2(\bar M^n)$ for all $n\in\overline{\N}_{\geq n_0}$. Moreover, since there exists $b\in(0,1)$ such that $\sup_{n\geq n_0}\|(\Delta K^n)^*_T\|_{L^\infty(P)}\leq b$, the process $\frac{1}{\E(-\beta  K^n)}=\frac{1}{\prod_{0<s\leq \cdot}(1-\beta\Delta K^n_s)}$ is increasing such that $\frac{1}{\E(-\beta K^n)}\geq 1$ and
\begin{align*}
\left\|\sup_{0\leq s\leq T}\left|\frac{1}{\E(-\beta K^n)_s}\right|\right\|_{L^\infty(P)}&\leq \left\|\exp\left(\sum_{0<s\leq T}-\beta \log(1-\beta\Delta K^n_s)\right)\right\|_{L^\infty(P)}\\
&\leq \exp\left(\frac{\beta}{1-\beta b}\sup_{n\geq n_0}\|K^n_T\|_{L^\infty(P)}\right)<\infty
\end{align*}
for all $n\geq n_0$ and any $\beta\in(0,\frac{1}{b})$. Since $K^\infty$ is of finite variation, both parts of the decomposition $K^\infty= \sum \Delta K^\infty+(K^\infty -\sum\Delta K^\infty)$ exist. Therefore we obtain by the estimates $1\leq \frac{1}{\E(-\beta K^\infty)}=\frac{1}{\E(-\beta \sum\Delta K^\infty- \beta(K^\infty-\sum\Delta K^\infty))}\leq e^{\left(\frac{\beta}{1-\beta b}+\beta\right)\|K^\infty_T\|_{L^\infty(P)}}$ that the increasing process $\frac{1}{\E(-\beta K^\infty)}$ is uniformly bounded and
$$\frac{1}{k}\|\vt\|_{L^2(\bar M^n)}\leq \|\vt\|_{\beta,n}\leq k\|\vt\|_{L^2(\bar M^n)}$$
holds with $k=\max\left( \exp\left(\frac{\beta}{1-\beta b}\sup_{n\geq n_0}\|K^n_T\|_{L^\infty(P)}\right), \left\|\frac{1}{\E(-\beta K^\infty)}_T\right\|_{L^\infty(P)}\right)$ for all $\vt\in\Theta_{S^n}$, for all $n\in\overline{\N}_{\geq n_0}$.

2) Following the remark after the proof of Corollary 5 in \cite{PRS98}, we apply Proposition 1 in \cite{PRS98} with $\beta>\mu^2>1$, $\vt=\vt^1-\vt^2$, $\psi=J^n(\vt^1)-J^n(\vt^2)$, $V_0=H_0^n(\vt^1)-H_0^n(\vt^2)$, $L=L^n\big(\bar M^n, H(\vt^1)\big)-L^n\big(\bar M^n,H(\vt^2)\big)$ and $C=\frac{1}{\E(-\beta K^n)}$ which gives that
\begin{align*}
\|J^n(\vt^1)-J^n(\vt^2)\|_{\beta,n}^2&=E\left[\int_0^T\frac{1}{\E(-\beta K^n)_s}\psi_s d\la \bar M^n\ra_s\psi_s \right]\\
&\leq \frac{1}{\mu^2}E\left[\int_0^T\frac{1}{\E(-\beta K^n)_s}\vt_s d\la \bar M^n\ra_s\vt_s \right]\\
&=\frac{1}{\mu^2}\|\vt^1-\vt^2\|_{\beta,n}^2,
\end{align*}
and therefore that $J^n$ is a contraction on $(\Theta_{S^n},\|.\|_{\beta, n})$ with $c:=\frac{1}{\mu^2}$ as modulus of contraction for all $n\in\overline{\N}_{\geq n_0}$.

3) This is an immediate consequence of 2) and Banach's fixed point theorem.
\ep
By part 3) of Lemma \ref{lJ} each F\"ollmer--Schweizer decomposition can be obtained for sufficiently large $n$ by a fixed point iteration in $p$. Then the next proposition says that these fixed point iterations converge for $p\to\infty$ even uniformly in $n$.
\begin{prop}\label{p1}
Under the assumptions of Theorem \ref{mthmconv}, there exists $n_0\in\N$ such that
$$\sup_{n\in \overline{\N}_{\geq n_0}}\|\xi^{n,p}-\widehat \xi^n\|_{L^2(M)}\overset{p\to\infty}{\longrightarrow}0.$$
\end{prop}
\bp
Using that there exist $n_0\in\N$ and $b\in(0,1)$ by Lemma \ref{lJ} such that the $J^n$ are contractions on $(\Theta_{S^n},\|.\|_{\beta, n})$ with a common modulus of contraction $c\in(0,1)$ independent of $n$, for any $\beta\in(1,\frac{1}{b})$, and that $\xi^{n,0}=0$ for each $n\in \overline{\N}_{\geq n_0}$, we obtain that
\begin{align}
\sup_{n\in \overline{\N}_{\geq n_0}}\big\|\xi^{n,p}-\widehat \xi^n\big\|_{L^2(\bar M^n)}&\leq k\sup_{n\in \overline{\N}_{\geq n_0}}\big\|\xi^{n,p}-\widehat \xi^n\big\|_{\beta,n}\nonumber\\
&\leq k c^p\sup_{n\in \overline{\N}_{\geq n_0}}\big\|\widehat\xi^{n}\big\|_{\beta,n}\leq k^2 c^p\sup_{n\in \overline{\N}_{\geq n_0}}\big\|\widehat\xi^{n}\big\|_{L^2(\bar M^n)}\label{pp1}.
\end{align}
To get an estimate for the right-hand side of \eqref{pp1}, we are going to use the continuity of the F\"ollmer--Schweizer decomposition and results on the equivalence of norms for \mbox{$\E$-local} martingales. To that end, we view each $S^n$ on $(\Om,\F,\FF^n,P)$. There we have that \mbox{$S^n=S_0+\bar M^n + \lambda^n\sint \la \bar{M}^n \ra$} is an $\E(-\lambda^n\sint \bar{M}^n)$-martingale (recall Definition \ref{defEmart}) by Corollary 3.17 in \cite{CKS98}, and $\E(-\lambda^n\sint \bar{M}^n)$ is regular and satisfies $R_2(P)$ with the same constant $\exp\left(\sup_{n\in \overline{\N}_{\geq n_0}}\|K^n_T\|_{L^\infty(P)}\right)$ for each $n\in \overline{\N}_{\geq n_0}$ by Proposition 3.7 in \cite{CKS98}. Therefore $S^n$ admits a F\"ollmer--Schweizer decomposition by and in the sense of Theorem 5.5 in \cite{CKS98}, which implies that $\|\widehat \xi^n\sint S^n_T\|_{\LiiP}\leq\| H^n\|_{\LiiP}$ for all $n\in\overline{\N}_{\geq n_0}$. As the constant in $R_2(P)$ is the same for all $n\in \overline{\N}_{\geq n_0}$, an inspection of the proof of Theorem 4.9 in \cite{CKS98} yields that 
$$\big\|\widehat \xi^n\sint S^n\big\|_{\cH^2(\FF^n)}\leq \bar c \big\|\widehat \xi^n\sint S^n_T\big\|_{L^2(P)}$$
also holds with the same constant $\bar c>0$ for all $n\in \overline{\N}_{\geq n_0}$, which implies
\be
\sup_{n\in \overline{\N}_{\geq n_0}}\big\|\widehat \xi^n\big\|_{L^2(\bar{M}^n)}\leq \sup_{n\in \overline{\N}_{\geq n_0}}\big\|\widehat \xi^n\sint S^n\big\|_{\cH^2(\FF^n)}\leq \bar c\sup_{n\in \overline{\N}_{\geq n_0}}\| H^n\|_{\LiiP}.\label{pp2}
\ee
Moreover, as $\frac{d\la \bar{M}^{n}\ra}{d\la M^{n}\ra}\to 1$ in $L^\infty(P_M)$ by our assumptions and part 1) of Lemma \ref{lKn}, there exists a constant $\tilde c>0$ such that
$$\frac{1}{\tilde c}\|\vt\|_{L^2(\bar{M}^n)}\leq\|\vt\|_{L^2(M)}\leq \tilde c\|\vt\|_{L^2(\bar{M}^n)}$$
for all $\vt\in\Theta_{S^n}=L^2(\bar{M}^n)$ and all $n\in \overline{\N}_{\geq n_0}$ by possibly enlarging $n_0$. Combining this with \eqref{pp1} and \eqref{pp2} gives that
$$\sup_{n\in \overline{\N}_{\geq n_0}}\big\|\xi^{n,p}-\widehat \xi^n\big\|_{L^2(M)}\leq k^2 c^p \bar c \tilde c \sup_{n\in \overline{\N}_{\geq n_0}}\big\|H^n\big\|_{L^2(P)}\overset{p\to\infty}{\longrightarrow}0,$$
since $\sup_{n\in \overline{\N}_{\geq n_0}}\big\|H^n\big\|_{L^2(P)}$ is bounded because $H^n\to H$ in $L^2(P)$. This completes the proof.
\ep
Before we can conclude the proof of Theorem \ref{mthmconv}, we need to establish not only the convergence of the fixed point iterations as the number of iterations $p$ tends to infinity, but also at each step as the mesh of the partitions goes to $0$.
\begin{prop}\label{p2}
Under the assumptions of Theorem \ref{mthmconv},
\be\label{p2eq}
\|\xi^{n,p}-\xi^{\infty,p}\|_{L^2(M)}\overset{n\to\infty}{\longrightarrow}0
\ee
for each $p\in\N_0$.
\end{prop}
\bp
We prove this by induction on $p\in\N_0$. To that end, we observe that \eqref{p2eq} is clearly true for $p=0$, as we have $\xi^{n,0}=\xi^{\infty,0}=0$, and so we assume as induction hypothesis that \eqref{p2eq} holds for $p\in\N_0$. By Lemma \ref{lconvMtoA} this implies that
$$H^{n,p}:=H^n-\int_0^T\xi^{n,p}_ud\bar{A}^n_u\longrightarrow H^{\infty,p}:=H-\int_0^T\xi^{\infty,p}_udA_u$$
in $\LiiP$ as $n\to\infty$. For each $n\geq n_0$ we can write
\begin{align}
\xi^{n,p+1}_t&=\xi_t(\bar{M}^n,H^{n,p})=\frac{E\left[H^{n,p}\Delta\bar{M}^n_{t_i}\big|\cF_{t_{i-1}}\right]}{E\left[(\Delta\bar{M}^n_{t_i})^2\big|\cF_{t_{i-1}}\right]}\nonumber\\
&=\left(\frac{E\big[H^{n,p}\Delta M^n_{t_i}\big|\cF_{t_{i-1}}\big]}{E\big[(\Delta M^n_{t_i})^2\big|\cF_{t_{i-1}}\big]}+\frac{E\big[H^{n,p}\Delta M^{A,n}_{t_i}\big|\cF_{t_{i-1}}\big]}{E\big[(\Delta M^{A,n}_{t_i})^2\big|\cF_{t_{i-1}}\big]}\cdot \frac{E\big[(\Delta M^{A,n}_{t_i})^2\big|\cF_{t_{i-1}}\big]}{E\big[(\Delta M^n_{t_i})^2\big|\cF_{t_{i-1}}\big]}\right)\frac{\Delta \la M^n\ra_{t_i}}{\Delta\la\bar{M}^n\ra_{t_i}}\nonumber\\
&=\left(\xi_t(M^n,H^{n,p})+\xi_t(M^{A,n},H^{n,p})\left(\frac{d\la M^{A,n}\ra}{d\la M^n\ra}\right)_t\right)\left(\frac{d\la M^n\ra}{d\la\bar{M}^n\ra}\right)_t\label{pp2eq1}
\end{align}
for $t\in[t_{i},t_{i+1})$ by plugging in $\bar{M}^n=M^n+M^{A,n}$ and the definition of the discrete-time GKW decomposition. Since
$$\|\xi(M^n,H^{n,p})-\xi(M^n,H^{\infty,p})\|_{L^2(M)}\leq\|H^{n,p}-H^{\infty,p}\|_{\LiiP}\to 0\qquad\text{as $n\to\infty$}$$
by the orthogonality of the terms in the GKW decomposition and
$$\xi(M^n,H^{\infty,p})\to\xi(M,H^{\infty,p})=\xi^{\infty,p+1}\qquad\text{as $n\to\infty$}$$
in $L^2(M)$ by Theorem 3.1 in \cite{JMP01}, we obtain that
\be
\xi(M^n,H^{n,p})\to\xi^{\infty,p+1}\qquad\text{as $n\to\infty$}\label{pp2eq2}
\ee
in $L^2(M)$. Moreover,
\begin{align}
\left\|\xi(M^{A,n},H^{n,p})\frac{d\la M^{A,n}\ra}{d\la M^n\ra}\right\|_{L^2(M)}&\leq\big\|\xi(M^{A,n},H^{n,p})\big\|_{L^2(M^{A,n})}\left\|\sqrt{\frac{d\la M^{A,n}\ra}{d\la M^n\ra}}\right\|_{L^\infty(P_M)}\nonumber\\
&\leq\big\|H^{n,p}\big\|_{\LiiP}\left\|\sqrt{\frac{d\la M^{A,n}\ra}{d\la M^n\ra}}\right\|_{L^\infty(P_M)}\longrightarrow 0\label{pp2eq3}
\end{align}
as $n\to\infty$ by our assumptions. Since these also give via part 1) of Lemma \ref{lKn} that $\frac{d\la M^n\ra}{d\la\bar{M}^n\ra}\to 1$ in $L^\infty(P_M)$, combining \eqref{pp2eq1}--\eqref{pp2eq3} implies that
$$\xi^{n,p+1}\longrightarrow\xi^{\infty,p+1}\qquad\text{as $n\to\infty$}$$
in $L^2(M)$, which completes the proof.
\ep
Now we have everything in place to finish the proof of Theorem \ref{mthmconv}.
\bp[Proof of Theorem \ref{mthmconv}]
The only remaining point is to show that we can control each of the terms in the decomposition
$$\widehat \xi^n-\widehat\xi=(\widehat \xi^n-\xi^{n,p})+(\xi^{n,p}-\xi^{\infty,p})+(\xi^{\infty,p}-\widehat\xi)$$
in a sufficient way. To that end, fix an arbitrary $\ve>0$. Then we choose $n_0$ and $p$ in $\N$ such that
$$\sup_{n\geq n_0}\|\xi^{n,p}-\widehat \xi^n\|_{L^2(M)}\leq \ve\qquad\text{and}\qquad\|\xi^{\infty,p}-\widehat\xi\|_{L^2(M)}\leq \ve$$
by Lemma \ref{lJ} and Proposition \ref{p1}. By possibly enlarging $n_0$, Proposition \ref{p2} allows us to obtain that
$$\|\xi^{n,p}-\xi^{\infty,p}\|_{L^2(M)}\leq \ve$$
for all $n\geq n_0$ and therefore that
$$\|\widehat\xi^{n}-\widehat\xi\|_{L^2(M)}\leq \sup_{n\geq n_0}\|\xi^{n,p}-\widehat \xi^n\|_{L^2(M)}+\|\xi^{n,p}-\xi^{\infty,p}\|_{L^2(M)}+\|\xi^{\infty,p}-\widehat\xi\|_{L^2(M)}\leq 3\ve,$$
which completes the proof.
\ep
Combining the previous results then gives the convergence of the LMVE strategies.
\bt
Suppose that $K$ is bounded, $\frac{d\la M^{A,n}\ra}{d\la M^{n}\ra}\overset{L^\infty(P_M)}{\longrightarrow}0$ and that there exist $n_0\in\N$ and $b\in(0,1)$ such that $\sup_{n\geq n_0}\|K^n_T\|_{L^\infty(P)}<\infty$ and $\sup_{n\geq n_0}\|(\Delta K^n)^*_T\|_{L^\infty(P)}\leq b$. Let $(\tau_n)_{n\in\N}$ be an increasing sequence of partitions of $[0,T]$ and $\hvt^n$ be the LMVE strategy with respect to $S^n$ on $(\Om,\F,\FF^n,P)$ and $\hvt$ the LMVE strategy with respect to $S$ on $(\Om,\F,\FF,P)$. Then $\hvt^n$ converges to $\hvt$ in $L^2(M)$ as $|\tau_n|\to 0$.
\et
\bp
Since $K^n=\la \lambda^n\sint M^n\ra^{\FF^n}$ and $K=\la \lambda\sint M\ra$ are bounded, $\E(\lambda^n\sint M^n)$ and $\E(\lambda\sint M)$ satisfy $R_2(P)$ and are regular with respect to $\FF^n$ and $\FF$, respectively, by Proposition 3.7 in \cite{CKS98}. By Corollary \ref{ces} this implies that $\hvt^n$ and $\hvt$ exist and are given by $\hvt^n=\frac{1}{\gamma}(\lambda^n-\widehat\xi^n)$ and $\hvt=\frac{1}{\gamma}(\lambda-\widehat\xi)$, where $\widehat\xi^n$ and $\widehat\xi$ denote the integrand of the F\"ollmer--Schweizer decomposition of $K^n_T$ and $K_T$. Since $K=\la \lambda\sint M\ra$ is bounded and hence $ \lambda\sint M$ is in $bmo_2$, the convergence of $\hvt^n$ to $ \hvt$ in $L^2(M)$ follows by combining Lemma \ref{lconvMtoA}, Corollary \ref{corKn} and Theorem \ref{mthmconv}, which completes the proof.
\ep
\appendix
\section{Representative square-integrable portfolios}
In this appendix we show the existence of representative square-integrable portfolios as announced in Section \ref{se:fp}. As stated in Lemma \ref{chap5:lrsip} below, these are strategies $\vp^i\in\TS_S$ for \mbox{$i=1,\ldots,d$}, which are representative in the sense that the financial market $(\widetilde S, \Theta_{\widetilde S})$ with \mbox{$\widetilde S^i:=\vp^ i\sint S$} for $i=1,\ldots,d$ generates the same wealth processes as the financial market $( S, \TS_S)$, i.e. \mbox{$\Theta_S\sint S=\Theta_{\widetilde S}\sint {\widetilde S}$}.  For this we use the notion of $\sigma$-square-integrability: A semimartingale $X$ is \emph{$\sigma$-square-integrable}, which we denote by $X\in\cH^2_{\sigma}(P),$ if there exists an increasing sequence $(D_n)$ of predictable sets such that $D_n\uparrow \OmT$ and \mbox{$\one_{D_n}\sint X\in\cH^2(P)$} for each $n$; see \cite{K04} for the concept of $\sigma$-localisation. If there exists a sequence of stopping times $(\sigma_n)$ such that we can choose $D_n=[\mskip-2mu[0,\sigma_n]\mskip-2mu]$ for each $n\in\N$, the concept of $\sigma$-square-integrability coincides with the classical notion of local square-integrability. The latter is for example always the case, if $S$ is continuous. The basic idea for the proof is then the following. Even though square-integrability is a global property of the strategy $\vt$ it implies that $\vt$ is $\sigma$-square-integrable, i.e.~$\vt\sint S\in\cH^2_{\sigma}(P)$, which can be characterised $\omt$-pointwise. Since there exists a one-to-one correspondence between $\sigma$-square-integrable and square-integrable integrands by Proposition 2 in \cite{E80} (see below), the $\omt$-pointwise characterisation of $\sigma$-square-integrability is sufficient to find the representative square-integrable portfolios. To derive this characterisation we need to work with the notion of predictable characteristics which we introduce next.

As in \cite{JS}, Theorem II.2.34, each semimartingale $S$ has the \emph{canonical representation}
$$S=S_0+S^c+\widetilde A +[x \mathbbm{1}_{\{|x|\leq 1\}}]\ast(\mu-\nu)+ [x \mathbbm{1}_{\{|x|> 1\}}]\ast\mu$$
with the jump measure $\mu$ of $S$ and its predictable compensator $\nu$. Then the quadruple $(b,c,F,B)$ of \emph{predictable characteristics} of $S$ consists of a predictable $\R^d$-valued process $b$, a predictable nonnegative-definite symmetric matrix-valued process $c$, a predictable process $F$ with values in the set of L\'evy measures and a predictable non-decreasing process $B$ null at zero such that
\be
\widetilde A= b\sint B, \qquad [S^c,S^c]=c\sint  B\qquad\text{and}\qquad\nu=F\sint B.\label{chap5:dc}
\ee
Using this local description of the semimartingale $S$ we can prove the existence of representative square-integrable portfolios.
\bl\label{chap5:lrsip}
There exist strategies $\vp^i\in\TS_S$ for $i=1,\ldots, d$ such that the financial markets $(S,\TS_S)$ and $(\widetilde S,\Theta_{\widetilde S})$ with $\widetilde S^i=\vp^i\sint S$ for $i=1,\ldots, d$ admit the same wealth processes, i.e.~$\Theta_S\sint S=\Theta_{\widetilde S}\sint \widetilde S$. 
\el
\bp 
 By Proposition 2 in \cite{E80} (and the paragraph preceding that), $\sigma$-square\--integrablity of a semimartingale $X$ is equivalent to the existence of a strictly positive, bounded predictable process $\psi$ such that $\psi\sint X\in\cH^2(P)$. As $\psi$ is bounded and strictly positive, we can therefore always switch back and forth between $\sigma$-square-integrable $X$ and square-integrable semimartingales $Y$ by using the associativity of the stochastic integral, i.e.~$Y=\psi\sint X$ and \mbox{$X=\frac{1}{\psi}\sint (\psi\sint X)=\frac{1}{\psi}\sint Y$}. Moreover, this also allows to reduce our problem to $\sigma$-square-integrability, which we consider first. Like any semimartingale, a stochastic integral $\vt\sint S$ of an $S$-integrable process $\vt$ is $\sigma$-square-integrable if and only if the sum of its squared jumps, $Z:=\sum_{0<s\leq \cdot}(\vt_s^\T\Delta S_s)^2$, is $\sigma$-integrable, i.e.~there exists an increasing sequence $(D_n)$ of predictable sets such that $D_n\uparrow \OmT$ and \mbox{$Z^n:=\one_{D_n}\sint Z$} has integrable total variation $\int_0^T|dZ^n_s|$ for each $n$.  By Theorem II.1.8 in \cite{JS}, the latter condition is equivalent to $\int_0^{\cdot}\int_{\R^d}(\vt_s^\T x)^2F_s(dx)dB_s$ being $\sigma$-integrable, which holds if and only if $\int_{\R^d}(\vt_s^\T x)^2F_s(dx)<+\infty$ $P_B$-a.e. If $S$ is one dimensional, i.e.~$d=1$, we can write $\vt_s^2\int_{\R^d}x^2F_s(dx)=\int_{\R^d}(\vt_s^\T x)^2F_s(dx)<+\infty$ $P_B$-a.e., which basically tells us that we must have $\vt=0$ $P_B$-a.e.~on the set  $D^c:=\{\int_{\R^d}x^2F(dx)=+\infty\}\in\cP$. Therefore setting $\vp^1:=\psi \mathbbm{1}_D$, where $\psi$ is the integrand from Proposition 2 in \cite{E80}  for the $\sigma$-square-integrable semimartingale $\mathbbm{1}_D\sint S$, gives the desired strategy.

In the multidimensional case, the situation is more involved due to the linear dependence between the different components of $S$. To deal with this issue, we use similar techniques as in \cite{CS09}, where we also refer the reader to for more explanations on problems arising from this. For the rest of the proof, we consider integrands $\vt\in L(S)$ as elements of \mbox{$L^0(\OmT,\cP,P_B;\R^d)$} and define the linear subspace $V$ by
$$V=\left\{\vt\in L^0(\OmT,\cP,P_B;\R^d)~\bigg|~\int_{\R^d}(\vt^\T x)^2F(dx)<+\infty \quad P_B\text{-a.e.}\right\}.$$
By definition, $V$ satisfies the stability property that $\vt^1\mathbbm{1}_D+\vt^2\mathbbm{1}_{D^c}\in 
V$ for all $\vt^1, \vt^2\in V$ and $D\in\cP$, and it is closed with respect to convergence in $P_B$-measure by Fatou's lemma. So there exist by Lemma 6.2.1 in \cite{DSbook} (see also Lemma 5.2 in \cite{CS09}) $v^i\in V$ for $i=1,\ldots, d$ such that
\bi
\item[\bf{1)}] $\{v^{i+1}\ne 0\}\subseteq\{v^i\ne 0\}$ for $i=1,\ldots,d-1$,
\item[\bf{2)}] $|v^i(\om,t)|=1$ or $|v^i(\om,t)|=0$,
\item[\bf{3)}] $(v^i)^\T v^k=0$ for $i\ne k$,
\item[\bf{4)}] $\vt\in V$ if and only if $\vt=\sum_{i=1}^d (\vt^\T v^i) v^i$ $P_B$-a.e.
\ei
Since $v^i$ is in $V$ and bounded by 2), $v^i\in L(S)$ and $v^i\sint S$ is $\sigma$-square-integrable for $i=1,\ldots,d$. By Proposition 2 in \cite{E80}, there exist strictly positive, bounded predictable processes $\psi^i$ such that $(\psi^iv^i)\sint S\in\cH^2(P)$ for $i=1,\ldots,d$, and we set $\vp^i=\psi^iv^i$ and \mbox{$\widetilde S^i=\vp^i\sint S$}. Since we can write each $\vt\in\TS_S\subseteq V$ as \mbox{$\vt=\sum_{i=1}^d (\vt^\T v^i) v^i=\sum_{i=1}^d \frac{(\vt^\T v^i)}{\psi^i} \vp^i$} $P_B$-a.e.~by 4), this gives $\widetilde \vt=(\frac{(\vt^\T v^1)}{\psi^1},\ldots,\frac{(\vt^\T v^d)}{\psi^d})=:\Psi \vt\in\Theta_{\widetilde S}$, where $\Psi:=\left(\frac{v^1}{\psi^1},\ldots,\frac{v^d}{\psi^d}\right)^\T$ is an $\R^{d\times d}$-valued predictable process, and that $\vt\sint S=\widetilde\vt\sint\widetilde S$ by the associativity of the stochastic integral. Conversely, we have for each $\widetilde \vt\in\Theta_{\widetilde S}$ that $\vt=\sum_{i=1}^d \widetilde \vt^i \vp^i=\Phi\widetilde \vt\in\TS_S$ with $\vt\sint S=\widetilde \vt\sint \widetilde S$, where $\Phi:=\left(\vp^1,\ldots,\vp^d\right)$ is an $\R^{d\times d}$-valued predictable process, which allows us to conclude that \mbox{$\Theta_S\sint S=\Theta_{\widetilde S}\sint \widetilde S$} and completes the proof.
\ep
\begin{remark}\label{chap5:repport}
SAs an alternative to the proof above one can introduce a predictable correspondence $C$ by
$$C\omt:=\left\{y\in\R^d~\bigg|~\int_{\R^d}(y^\T x)^2F(dx)<+\infty\right\}$$
for all $\omt\in\OmT$. Then the condition $\vt\in V$ can be formulated as the pointwise constraint that $\vt\omt\in C\omt$ $P_B$-a.e. As the values of $C$ are linear subspaces, one can deduce the existence of representative $\sigma$-square-integrable portfolios by using (the arguments in the proof of) Theorem B.3 in Nutz \cite{N09}. The correspondence of the transformed constraints $\widetilde C$ is then of course equal to $\R^d$ for all $\omt\in\OmT$ and the representative $\sigma$-square-integrable portfolios are the representative portfolios.
\end{remark}
\begin{ak}
The author thanks Tahir Choulli, Michael Kupper and Martin Schweizer for discussions and Wolfgang Runggaldier, Martin Schweizer and two anonymous referees for careful reading and helpful suggestions. Financial support by the National Centre of Competence in Research ``Financial Valuation and Risk Management'' (NCCR FINRISK), Project D1 (Mathematical Methods in Financial Risk Management) is gratefully acknowledged. The NCCR FINRISK is a research instrument of the Swiss National Science Foundation.
\end{ak}

\end{document}